\documentclass[12pt,preprint]{aastex}

\shorttitle{NIR Knots in Cas A}
\shortauthors{Lee et al.}

\usepackage{amsmath}
\usepackage{amsfonts}
\usepackage{amssymb}

\newcommand{\msun}{M_{\sun}}
\newcommand{\kms}{{\rm km~s}^{-1}}

\newcommand{\ergscmsr}{{\rm erg}~{\rm s}^{-1}~{\rm cm}^{-2}~{\rm sr}^{-1}}
\newcommand{\ha}{H$\alpha$}
\newcommand{\hi}{H {\small I}}
\newcommand{\hii}{H {\small II}}
\newcommand{\hei}{He {\small I}}
\newcommand{\ci}{[\ion{C}{1}]}
\newcommand{\nn}{[\ion{N}{1}]}
\newcommand{\nii}{[\ion{N}{2}]}

\newcommand{\oii}{[\ion{O}{2}]}

\newcommand{\sili}{[\ion{Si}{1}]}

\newcommand{\silx}{[\ion{Si}{10}]}
\newcommand{\silvi}{[\ion{Si}{6}]}
\newcommand{\pii}{[\ion{P}{2}]}
\newcommand{\sii}{[\ion{S}{2}]}
\newcommand{\siii}{[\ion{S}{3}]}
\newcommand{\feii}{[\ion{Fe}{2}]}
\newcommand{\feiii}{[\ion{Fe}{3}]}

\begin{document}

\title{Near-Infrared Knots and Dense Fe Ejecta
in the Cassiopeia A Supernova Remnant}

\author{Yong-Hyun Lee\altaffilmark{1},
Bon-Chul Koo\altaffilmark{1},
Dae-Sik Moon\altaffilmark{2},
Michael G. Burton\altaffilmark{3,4},
Jae-Joon Lee\altaffilmark{5}}

\email{yhlee@astro.snu.ac.kr}
\altaffiltext{1}{Department of Physics and Astronomy, Seoul National University,
Seoul 151-742, Korea}
\altaffiltext{2}{Department of Astronomy and Astrophysics, University of Toronto,
Toronto ON M5S 3H4, Canada}
\altaffiltext{3}{School of Physics, University of New South Wales,
Sydney, NSW 2052, Australia}
\altaffiltext{4}{Armagh Observatory and Planetarium, College Hill, Armagh, BT61 9DG,
Northern Ireland, UK}
\altaffiltext{5}{Korea Astronomy and Space Science Institute,
Daejeon 305-348, Korea}

\begin{abstract}
We report the results of broadband (0.95--2.46~\micron) near-infrared
spectroscopic observations of the Cassiopeia A supernova remnant.
Using a clump-finding algorithm 
in two-dimensional dispersed images,
we identify 63 `knots' from eight slit positions and
derive their spectroscopic properties.
All of the knots emit \feii\ lines together with
other ionic forbidden lines of heavy elements,
and some of them also emit H and He lines.
We identify 46 emission line features in total from the 63 knots
and measure their fluxes and radial velocities.
The results of our analyses of the emission line features based on
principal component analysis show that
the knots can be classified into three groups:
(1) He-rich, (2) S-rich, and (3) Fe-rich knots.
The He-rich knots have relatively small, $\lesssim 200~{\rm km~s}^{-1}$,
line-of-sight speeds
and radiate strong \hei\ and \feii\ lines resembling closely
optical quasi-stationary flocculi of circumstellar medium,
while the S-rich knots show strong lines from O-burning material with
large radial velocities up to $\sim 2000~{\rm km~s}^{-1}$
indicating that they are supernova ejecta material known as
fast-moving knots.
The Fe-rich knots also have large radial velocities
but show no lines from O-burning material.
We discuss the origin of the Fe-rich knots and conclude that
{\em they are most likely ``pure'' Fe ejecta
synthesized in the innermost region during the supernova explosion.}
The comparison of \feii\ images with other waveband images shows that
these dense Fe ejecta are mainly distributed along the southwestern shell
just outside the unshocked $^{44}$Ti in the interior,
supporting the presence of unshocked Fe associated with $^{44}$Ti.
\end{abstract}

\keywords{infrared: ISM --- ISM: individual objects (Cassiopeia A) ---
ISM: supernova remnants --- line: identification --- methods: statistical}

\section{Introduction} \label{sec-int}
A massive star builds up onion-like layers of different chemical elements
synthesized by hydrostatic nuclear burning processes during its lifetime.
At the end of its evolution,
the innermost Fe core collapses into a neutron star,
which triggers a core-collapse supernova (SN) explosion.
The detailed process of the explosion is complicated and poorly understood,
but a consensus from theoretical studies suggests that 
the explosion should be asymmetric and turbulent, especially near the core
\citep[e.g.,][and references therein]{sum05,tak14,gil15}.
Multi-dimensional numerical simulations have shown that 
the explosion also leads to extensive mixing and inversion
among the stratified layers by hydrodynamic instabilities
\citep[e.g.][]{kif03,kif06,ham10,mao15}.
A distinct method to explore the explosion dynamics of SNe, therefore,  
would be to investigate the detailed chemical and kinematic properties 
of SN ejecta material in nearby young Galactic supernova remnants (SNRs)
where the imprints of explosion remain.

Cassiopeia A (hereafter Cas A),
at the age of $\sim 340$ years \citep{tho01,fes06b},
is one of the best studied young Galactic SNRs.
Its SN explosion was classified as Type IIb
based on the optical spectra of light echoes
\citep{kra08,res11},
which implies that the progenitor was probably a star of $15$--$20~\msun$
that had lost a significant portion, but not all,
of its H-rich envelope before the explosion.
Over the past several decades, Cas~A has been extensively studied in 
almost all wavebands from radio to gamma-rays.
The complex spatial distribution of {\em shocked} SN ejecta, such as
the X-ray-emitting Fe-rich ejecta plumes beyond the Si-rich material
and O/S-rich optical knots outlying the bright ejecta shell
\citep[e.g.,][]{hug00, ham08},
indicates an explosion resulting in an inversion of the chemical layers.
Furthermore, the inhomogeneous distribution of {\em unshocked} SN ejecta
radiating Si and Ti emission lines \citep{ise10,gre14} implies that
the explosion was turbulent near the progenitor core.

In the visible waveband,
many bright optical knots are presented in and around Cas A.
They have been classified into two major groups based on
their proper motions and line-of-sight velocities:
(1) fast-moving knots (FMKs) and (2) quasi-stationary flocculi (QSFs).
The FMKs show large proper motions and high radial velocities,
corresponding to expansion velocities of up to $\sim10^{4}~\kms$
\citep[e.g.,][]{fes96,ham08}.
They have spectral features strongly enhanced in
O and other heavy elements (e.g., S and Ar)
that are mostly synthesized from the nuclear burning process
in a deep stellar layer,
while showing no detectable H or He emission lines
\citep[e.g.,][]{fes96,hur96}.
Based on their high expansion velocities and chemical composition,
they have been regarded as dense material ejected from
the disrupted layer of the progenitor after the SN explosion.
The dynamical and chemical properties of the QSFs are very different from
those of the FMKs.
The QSFs have considerably slower velocities ($v<500~\kms$), and so
their typical expansion age is $11{,}000\pm2000$ years \citep{van85}, 
which is much larger than the age of the remnant.
They are bright in
\nii\ and \ha\ emission,
with a handful of other H and He emission lines,
in optical spectra \citep[e.g.,][]{hur96,ala14}.
Considering these properties,
QSFs are believed to be dense circumstellar medium (CSM)
blown out from the progenitor prior to the SN explosion.
In addition, some intermediate optical knots,
that is, the so called fast-moving flocculi (FMFs) or nitrogen knots (NKs),
were reported by \citet{fes87}.
While their spectra, which show strong
\nii\ 6548, 6583 \AA\footnote{All wavelengths listed in this paper are ``air wavelengths.''}
accompanied by weak \ha\ without any lines of O and S,
are analogous to those of QSFs \citep{fes87,fes06a,fes01a},
their proper motions are larger than $0\farcs3~{\rm yr}^{-1}$,
corresponding to $\sim 5000~\kms$ \citep{ham08}.
Most of them have been found outside of the bright main ejecta shell.
Therefore, these outlying N-rich knots have been interpreted as
being fragments of the progenitor's photosphere
expelled by the SN blast wave at the time of explosion.

The dynamical and chemical properties of these different types of optical knots
have provided important clues to unveiling the SNR's origin and evolution.
For example, the expansion center and age were determined from
the proper motions of the FMKs
\citep[e.g.,][and references therein]{tho01,fes06b},
and the three-dimensional (3D) structure of the ejecta knots
reconstructed from spectral mapping observations
has shown that the SN ejecta is expanding spherically
but is systematically receding at a speed of $800~\kms$
at a distance of 3.4 kpc \citep{ree95,del10,mil13,ala14}.
The dense, slow-moving QSFs indicate that 
the progenitor had undergone significant and inhomogeneous mass loss
during the red supergiant phase \citep{che03}.

Although this velocity-based classification is an efficient way
to classify the knots as SN ejecta and CSM,
which have distinctive expansion velocities,
i.e., a few $1000~\kms$ vs. a few $100~\kms$,
it encounters limitations when characterizing the SN ejecta material from
the different nucleosynthetic layers.
According to previous numerical simulations for core-collapse SN explosions,
the radial velocity profiles of heavy elements in SN ejecta
are almost identical \citep{kif06,jog09}.
Several models in \citet{jog09} predict
the velocity separation of up to $\sim1000~\kms$
between different heavy elements.
Their velocity profiles, however, are very broad (a few $1000~\kms$),
so that their distributions largely overlap.
These numerical simulations may imply that
SN material from different nucleosynthetic layers is barely distinguishable
in velocity space.
In order to comprehend the explosion dynamics, 
therefore, a more systematic classification of SN ejecta
based on their chemical composition is needed.

In this paper,
we report the results of broad near-infrared (NIR) spectroscopic observations
toward the main ejecta shell of Cas A,
focusing on classification of the emission knots
based on their {\em spectrochemical} properties.
The NIR study of Cas A has been relatively limited in the literature,
although there are many bright forbidden lines of various elements
in the NIR waveband,
some of which may arise from the deep nucleosynthetic layers.
As far as we are aware,
the only NIR spectroscopic study covering the entire {\it JHK} bandpass
was conducted by \citet{ger01},
who obtained low-resolution ($R \sim 700$) spectra
of five FMKs and three QSFs that were previously known.
They showed that the spectra of the FMKs are dominated by
\sii\ 1.029, 1.032, 1.034, 1.037 \micron\
(hereafter \sii\ 1.03 \micron\ multiplets)
and \pii\ 1.188 \micron,
as well as two high-ionization Si lines,
\silvi\ 1.963 \micron\ and \silx\ 1.430 \micron,
while those of the QSFs show strong \hei\ 1.083 \micron\
accompanied by H emission lines. 
A dozen bright \feii\ lines are also detected in both FMKs and QSFs.
Our spectra confirm these features.
Here we present the spectrochemical classification of knots 
and discuss their characteristics.
The organization of the paper is as follows. 
In Section~\ref{sec-obs},
we outline our spectroscopic observations and data reduction procedures.
An explanation of how we identified individual knots
from two-dimensional (2D) dispersed images
and derived their spectral properties
is given in Section~\ref{sec-ide}.
In Sections~\ref{sec-pca} and \ref{sec-dis}, 
we carry out a classification of the knots using
principal component analysis (PCA) and
discuss the origin of knots in different classes.
Finally, the paper is summarized in Section~\ref{sec-sum}.

\section{Observations and Data Reduction} \label{sec-obs}
\subsection{Near-infrared Spectroscopy} \label{sec-obs-spt}
\begin{figure}
\center{
\includegraphics[scale=0.9]{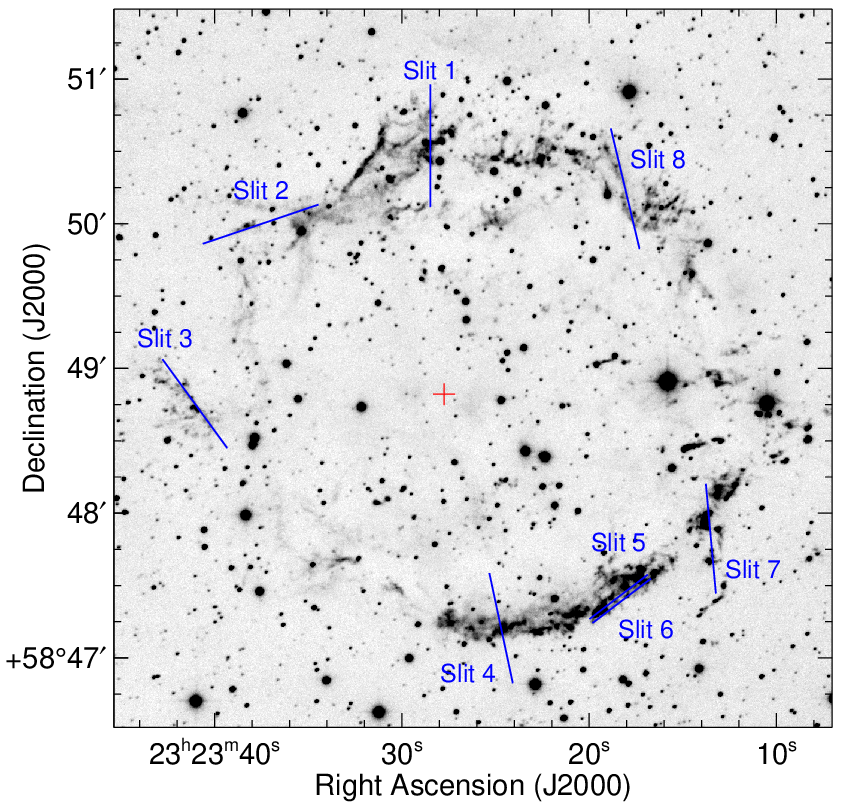}
}
\caption{
	Finding chart for the eight long-slit positions.
	The background is a \feii\ 1.644 \micron\ narrow-band image
	observed in 2008 August.
	The red cross mark is the expansion center
	($\alpha$ = 23$^{h}$23$^{m}$27$^{s}$.77 $\pm$ 0$^{s}$.05,
	$\delta$ = 58$^{\circ}$48$\arcmin$49$\farcs$4 $\pm$ 0$\farcs$4 [J2000])
	of the SN ejecta measured by \citet{tho01}.
	Note that the length of each blue bar represents
	the total effective slit coverage
	depending on the nodding offset (see Table~\ref{tbl-log}).
	North is up and east is to the left.
} \label{fig-slit}
\end{figure}

We carried out NIR spectroscopic observations of Cas A using TripleSpec
mounted on the Palomar 5 m Hale telescope.
TripleSpec is a cross-dispersed NIR spectrograph that
provides simultaneous wavelength coverage from 0.94 to 2.46 \micron\
at a spectral resolving power of $R \sim 2700$ \citep{wil04,her08}.
The spectrograph uses two adjacent quadrants,
i.e., $2048 \times 1024$ pixels, of a Rockwell Scientific Hawaii-II array.
The slit width and length are 1\arcsec\ and 30\arcsec, respectively.
On 2008 June 29 and August 8,
we obtained spectra at eight slit positions
along the main ejecta shell (Figure~\ref{fig-slit}).
The orientations of Slits 1 and 4 were set perpendicular to the shell,
while those of the others were largely parallel to the shell.
The nearby A0V star HD223386 was observed as a spectroscopic standard
right before and/or after the target observations at similar airmasses.
In addition to the target spectra,
we also obtained spectra of sky background by dithering along the slit
or taking spectra of nearby sky,
depending on the complexity of the target fields.
The total on-source exposure time at each slit position
ranges from 300 to 1800 s.
The detailed parameters of the spectroscopic observations are provided in
Table~\ref{tbl-log}.

We developed an IDL-based data reduction pipeline
to reduce the obtained TripleSpec data.
The pipeline first performed subtraction of a dark frame and flat fielding,
followed by subtraction of sky background emission.
For the latter, we used sky background emission
obtained in a dithered frame or in a frame from nearby empty sky,
depending on the complexity of the source emission in a given slit.
Because of the non-uniform dispersion by the three cross-dispersing prisms
in TripleSpec \citep{wil04,her08},
some orders of the TripleSpec spectra are severely curved.
By carefully tracing the spectra of standard stars and airglow emission lines,
we obtained a 2D wavelength solution for each order
that can correct this effect in a satisfactory manner.
We conducted fifth-order order polynomial fits
to the wavelengths of the OH airglow emission lines \citep{ost97,rou00}
in the TripleSpec spectra
to obtain the wavelength solutions at 0.5 \AA\ $1\sigma$ uncertainty
for each order.
Heliocentric velocity correction was also performed
when calculating the velocity of the emission line
detected in the TripleSpec spectra.
We used an A0V-type standard star (HD223386)
in the photometric calibration and confirmed that
the fluxes of the \feii\ 1.644 \micron\ emission line are consistent with
those from the narrow-band imaging observations of the same areas
(see Section~\ref{sec-obs-img}).

\subsection{Near-infrared Imaging} \label{sec-obs-img}
In 2005 August 28 and 2008 August 11,
we performed NIR imaging observations for the remnant
using the Wide-field Infrared Camera \citep[WIRC;][]{wil03}
attached to the Palomar 5 m telescope.
The camera consists of
a single $2048 \times 2048$ Rockwell Hawaii-II HgCdTe NIR detector
with a pixel scale of $0\farcs2487$ pixel$^{-1}$,
which provides a field of view of $8\farcm7 \times 8\farcm7$.
We used an \feii\ narrow-band filter
that has a mean wavelength of 1.644 \micron\ and a bandwidth of 252 \AA.
While the single exposure time per frame in 2005 and 2008 is 60 s and 200 s,
the multiple dithering observations yield total integration times of
1200 s and 5400 s, respectively.
We also obtained H-continuum narrow-band images
(mean wavelength of 1.570 \micron\ and bandwidth of 236 \AA)
in order to subtract the bright stars in the \feii\ narrow-band images.
The average seeing throughout our observations was $\sim 0\farcs9$ FWHM.
We followed standard procedures for the reduction of NIR imaging data.
First, the dark and sky background
were subtracted from each individual dithered frame.
Then, all of the frames were divided by the normalized flat image.
The astrometric solution was derived on the basis of
unsaturated stars around the remnant
in the Two Micron All Sky Survey (2MASS) Point-Source Catalog
\citep[PSC;][]{skr06}.
We coadded all dithered frames which were astrometrically aligned.
In terms of photometric calibration,
the {\it H}-band magnitude in the 2MASS system was used by assuming that
the \feii\ narrow-band magnitude is the same as the {\it H}-band magnitude.
The uncertainty in the zero-point magnitude we derived
is less than 0.1 mag, corresponding to the 10\% of the flux.

\section{Identification of Knots and Line Parameters} \label{sec-ide}
\begin{figure}
\center{
\includegraphics[scale=0.9]{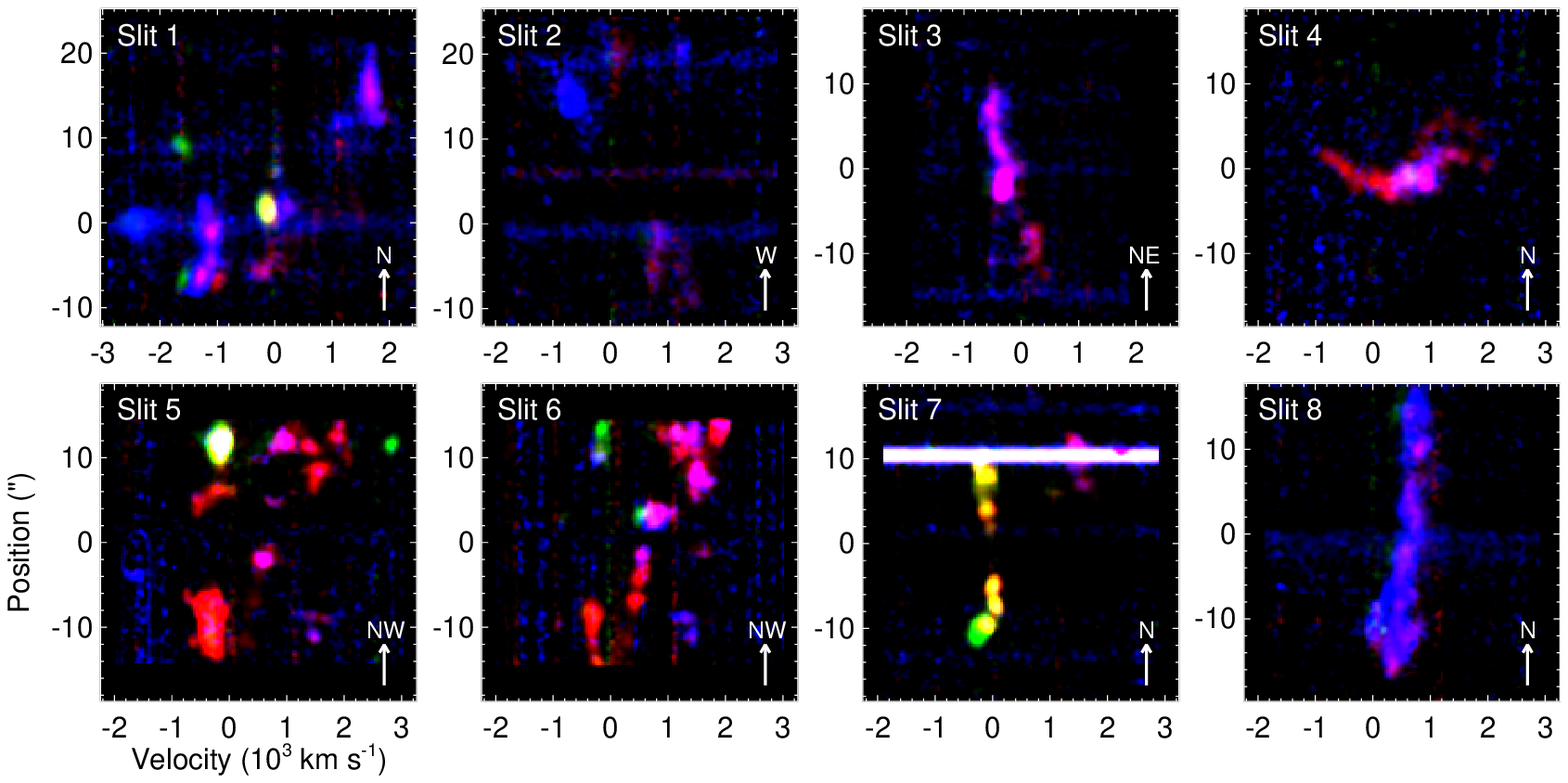}
}
\caption{
	Two-dimensional dispersed image of bright emission lines in three colors:
	\feii\ 1.644~\micron\ in red,
	\hei\ 1.083~\micron\ in green, and
	\pii\ 1.188~\micron\ + \siii\ 0.953~\micron\ in blue.
	The $y$-axis represents the position along the slit length
	with zero corresponding to the slit center in Table~\ref{tbl-log}.
	The dynamic ranges of RGB colors are the same,
	i.e., maximum to minimum intensity ratios of 10,
	while their maximum intensities have a ratio of 1:10:70 (R:G:B)
	in each slit image.
	Note that the continuous spectrum at $+10\arcsec$ in Slit 7
	is of an adjacent star falling on the slit position.
	The slit direction is marked in the lower right of each panel
	(see Table~\ref{tbl-log} for details).
} \label{fig-2dspec}
\end{figure}

\begin{figure}
\center{
\includegraphics[scale=0.9]{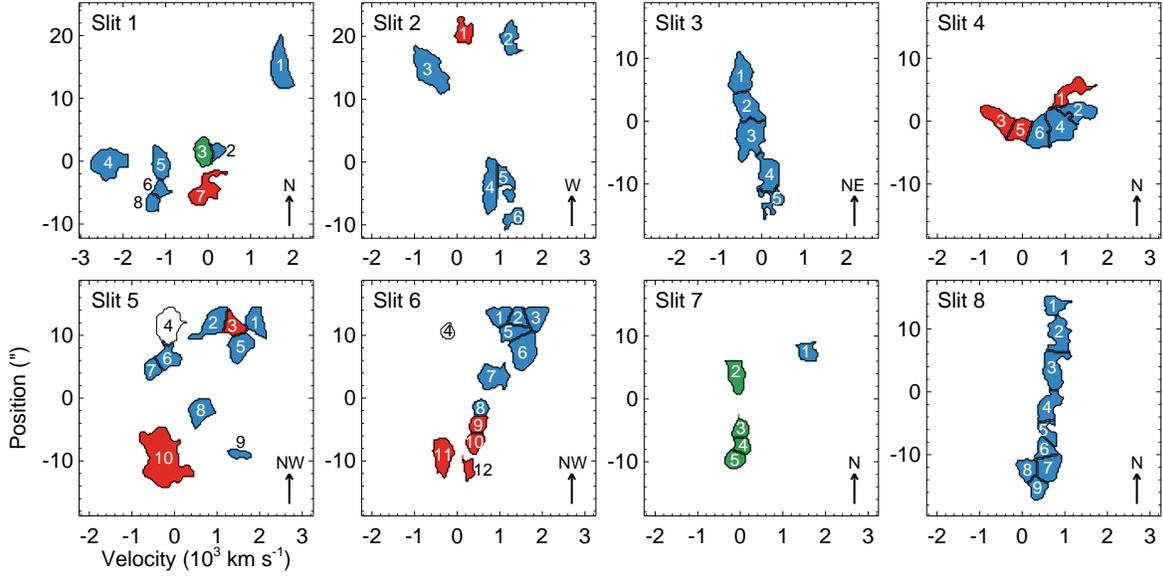}
}
\caption{
	63 knots identified by the {\sc Clumpfind} code
	in the 2D dispersed images.
	The colors represent the different types of knots:
	He-rich knots is in green, S-rich knots in blue,
	and Fe-rich knots in red.
	Knot 4 in Slits 5 and 6, which are not colored, are composed of
	two knots of different types with
	almost the same position and velocity (see Table~\ref{tbl-knot}).
	In Slit 7, there is another knot above Knot 2
	(see Figure~\ref{fig-2dspec}),
	but it spatially overlaps with a continuum source
	and was not identified as a clump by the {\sc Clumpfind} code.
	The slit direction is marked in the lower right of each panel
	(see Table~\ref{tbl-log} for details).
} \label{fig-clumps}
\end{figure}

Figure~\ref{fig-2dspec} shows 2D dispersed images of
strong emission lines detected in our TripleSpec spectra:
red for \feii\ 1.644 \micron, green for \hei\ 1.083 \micron,
and blue for \pii\ 1.188 \micron\ + \siii\ 0.953 \micron.
The emission features are complex, often with multiple peaks,
and so the identification of individual `knots' by visual inspection
is not straightforward.
We used the IDL routine {\sc Clumpfind} \citep{wil94},
which was developed for the identification of clumps in molecular clouds.
The routine identifies `clumps' by searching for local maxima
above some intensity threshold and
following them down them to low-intensity levels.
For a given 2D dispersed image,
a ``mask'' locating individual knots was generated from
\feii\ 1.644 \micron\ emission features or, if they are weak,
\siii\ 0.953 \micron\ or \pii\ 1.188 \micron\ emission features,
and shifted along the wavelength
to find other emission lines associated with the knots.
The details of this knot identification process are given in
the Supplement Material of \citet{koo13}.
In total, we identified 63 knots of
distinctive kinematical and spectral properties in the 2D dispersed images.
Figure~\ref{fig-clumps} shows their locations
and Table~\ref{tbl-knot} lists their positions, sizes, and radial velocities.

We extracted one-dimensional (1D) spectra of individual knots
using their mask files (Figure~\ref{fig-clumps}).
We identified 46 emission lines in total
and derived their parameters by performing single Gaussian fits
to the detected lines.
Table~\ref{tbl-flx} lists the derived line widths and fluxes.
\feii\ 1.644 \micron\ line is detected in all 63 knots
and their fluxes are also listed in Table~\ref{tbl-knot}.
Among the 46 emission lines,
43 lines are previously detected lines in SNRs
\citep{den81,rud94,hur96,ger01},
whereas three lines of \feiii\ at 2.145, 2.218, and 2.242 \micron\
are detected for the first time in SNRs.
The \feiii\ lines originate from transitions
between levels in $^3$G and $^3$H terms 
with high excitation energies ($\gtrsim 30{,}000$ K) and   
have been detected in a few objects, such as
SgrA$^{\ast}$/IRS16 complex \citep[][and references therein]{lut93},
classical novae \citep{wag96},
\hii\ regions \citep{oku01}, and planetary nebulae \citep{lik06}.
Among the previously reported lines, on the other hand,
the O {\small I} 1.1286, 1.1287 \micron, \feii\ 1.1881 \micron, and
\silx\ 1.430 \micron\ lines are not detected in our spectra.
\citet{ger01} reported detection of the
O {\small I} lines in the FMKs in Cas A,
but we were unable to confirm the detection with our spectra.
They also reported detection of
the highly ionized \silx\ 1.430 \micron\ line,
but again we were unable to confirm the detection.
The \feii\ 1.1881 \micron\ line was included in the list of identified lines
in three QSFs in Cas A and the Kepler SNR by \cite{ger01},
but we consider that
this was a misidentification of the \pii\ 1.1883 \micron\ line.  
The expected flux of the \feii\ 1.1881 \micron\ line in typical conditions
(e.g., $T \lesssim 10^{4}$ K, $n_{\rm e} \lesssim 10^{5}$ cm$^{-3}$) in SNRs
is almost negligible,
i.e., its flux relative to the \feii\ 1.257 \micron\ line is
$\lesssim 10^{-5}$,
whereas the \pii\ 1.1883 \micron\ line could be as strong as
$\sim 1/10$ of the \feii\ 1.257 \micron\ line for cosmic abundance,
or even higher if Fe atoms are locked into dust grains \citep{koo13}.

\section{Principal Component Analysis of Knots' Spectral Properties}
\label{sec-pca}
\subsection{Method} \label{sec-pca-method}
In order to systematically characterize the spectral properties of the 63 knots,
which have 46 emission lines in total, we applied the PCA method.
PCA measures the variances among the original input variables
(i.e., brightness of the emission lines in this study)
and then sets new orthogonal axes called principal components (PCs)
along the largest variances.
Therefore, the largest variance (or information) is contained in
the first PC (PC1), the second most in PC2, and so forth.
If there are significant correlations among the original input variables,
then the majority of the information is confined within the first few PCs,
which makes it possible to categorize the objects into
a few groups based on the first few PCs.

Before performing the PCA, we apply an extinction correction 
using the line ratio of the \feii\ 1.257 and 1.644 \micron\ emission.
These two lines originate from the same upper level ($a^{4}D_{7/2}$)
and therefore their intrinsic flux ratio
($[F_{\rm 1.257}/F_{\rm 1.644}]_{\rm int}$)
depends on the Einstein A-coefficients ($A_{\rm ki}$) and their wavelengths,
i.e., $[F_{\rm 1.257}/F_{\rm 1.644}]_{\rm int}=(A_{ki,1.257}/1.257)/(A_{ki,1.644}/1.644)$.
The A-coefficients, however, are considerably uncertain
so the line ratio ranges from 0.98 to 1.49
\citep[][and references therein]{gia15,koo15} in literature.
We adopt a line ratio of 1.36,
which is the value suggested by \citet{nus88} and \citet{deb10}.
(We found that the uncertainty of the theoretical line ratio
does not affect our classification of the knots
because they are well grouped in PC spaces,
as we will show in Section~\ref{sec-pca-result-pc}.
The criteria of the groups, however, may change
depending on the intrinsic line ratio we adopt.
In Section~\ref{sec-pca-result-pc}, we will describe this in more detail.)
Then, by comparing the observed line flux ratio to the intrinsic ratio,
we obtain the extinction of the knots
and deredden the observed fluxes of all of the lines.
We use the general interstellar extinction curve derived from
a carbonaceous-silicate grain model with a Milky Way size distribution
for $R_{V}=3.1$ \citep{dra03}.
The derived extinctions are listed in Table~\ref{tbl-knot}.
In our previous work \citep{leeyh15},
we showed that the extinction toward the west is systematically larger than
that toward the east,
which is consistent with the previous optical/X-ray extinction estimates
\citep[see Figure~1 in][]{leeyh15}.
We further showed that the extinctions of red-shifted knots are
systematically higher than those of the blue-shifted knots,
implying the presence of a large amount of SN dust inside
and around the main ejecta shell
\citep[see][for details]{leeyh15}.
The lines from the same upper level in the dereddened spectral data
do not provide independent information any more, 
so that the number of attributes in the PCA are now reduced from 46 to 23.
In order to prevent a few bright lines dominating the PCA,
the line intensities are standardized by
subtracting the mean and dividing by the standard deviation.
Since we use the mean-subtracted data,
the zero PCs represent the location of mean brightness,
not the location of the zero fluxes of the lines (hereafter convergent point),
which is important for our classification
(see Section~\ref{sec-pca-result-pc}).
In order to get the PC coefficients of a knot indicating the convergent point
when all the emission lines of a knot get close to zero flux,
we add one artificial knot into the data set
that has emission lines with zero flux.

\subsection{Results} \label{sec-pca-result}
\subsubsection{Principal Components and Classification of Knots}
\label{sec-pca-result-pc}
Table~\ref{tbl-pca} contains the relative and cumulative fraction of variances
contributed by the 10 most significant PCs.
The first three PCs account for the majority, i.e., $\sim$85\%,
of the spectral information;
thus, we use them in our classification of the knots.
Figure~\ref{fig-atr} shows the projection of the 23 attributes
on the combination plane of the three most significant PCs.
(This type of plots is known as an {\it h-plot};
see \citet{ung97} and \citet{neu07}.)
While Figure~\ref{fig-atr}(a) and \ref{fig-atr}(b) show 2D projections
on the planes of (PC1 vs. PC2) and (PC1 vs. PC3),
Figure~\ref{fig-atr}(c) shows 3D projections for
the three PCs together.
The lengths of the vectors in the plots are proportional to
the fractional contributions by the spectral line to the given PC,
and their quadratic sum is equal to unity.
In Figure~\ref{fig-atr}(d),
we visualize the 3D projections using the coordinate of
(Longitude vs. Latitude) representing the two projection angles
on the surface of a sphere.
We see in Figure~\ref{fig-atr} that
the attributes can be largely divided into three groups,
each of which is composed of several strongly correlated spectral lines. 
(Note that the cosine of the angle between the vectors on the plots
measures the linear correlation between the emission lines;
see \citet{neu07} for example.)
The first group (hereafter `He group') is composed of
\hi, \hei, and \nn\ lines.
These lines are almost perfectly correlated with each other.
The lengths of their vectors are close to unity in the PC1-PC3 plane,
which means that these spectral lines
are properly accounted for by PC1 and PC3.
The second group (`S group') is composed of
\silvi, \pii, and ionized S emission lines.
These lines are also strongly correlated with each other
and mostly contributed by PC2 and PC3
in the direction orthogonal to the He group lines.
The last group (`Fe group') is composed of ionized Fe emission lines,
i.e., \feii\ and \feiii\ lines.
These lines are rather loosely correlated
but are still generally well separated from the lines in the other two groups.
The \feii\ 2.046 and 2.224 \micron\ lines in particular appear
somewhat distinct from the other \feii\ lines. 
This might be caused by the higher excitation energies of the two lines
than those of the other lines, i.e., $\sim 30{,}000$ K vs. $\sim 10{,}000$ K.

We plot the PC coefficients of the 63 knots on the PC planes
in Figure~\ref{fig-obj}. 
The central positions of the planes, where (PC1, PC2, PC3) = (0, 0, 0),
represent the spectrum made by averaging all of the spectra of all 63 knots.
In the lower panels of the figure,
which are the enlarged plots of the central areas of the PC planes,
we draw dashed lines in order to group the knots (see below) originating from
(PC1, PC2, PC3) = (-0.10, 0.23, 0.10).
This convergent point is the location of the virtual knot with zero flux
(Section~\ref{sec-pca-method}),
and the radial distance from the convergent point is proportional to
the brightness of the emission line.
The distributions of the PC coefficients in Figure~\ref{fig-obj} are
very similar to those in Figure~\ref{fig-atr}.
There appear to be three groups of knots in Figure~\ref{fig-obj}
that have PC coefficients similar to those of the three groups
in Figure~\ref{fig-atr}, i.e., the He, S, and Fe groups.
We therefore group the knots in Figure~\ref{fig-obj} as
He-rich, S-rich, and Fe-rich knots using the dashed lines.
As a result,
we identify 7 He-rich knots, 45 S-rich knots, and 11 Fe-rich knots.

Figure \ref{fig-1dspec} shows sample 1D spectra of the three groups.
As expected, the He-rich knots have
strong lines of \hei\ 1.083 \micron\ compared to \feii\,
and some of them also show several emission lines of
\hi, \nn, \pii, and \feiii\ as well.
The \hi\ and \nn\ lines are detected only in the He-rich knots.
The S-rich knots also have bright \feii\ lines,
but they have much stronger lines of
\siii\ 0.953 \micron, \sii\ 1.03 \micron\ multiplets,
\pii\ 1.188 \micron, and \silvi\ 1.963 \micron.
The low-ionized emission lines of refractory elements, i.e., \ci\ and \sili,
are also detected in several S-rich knots.
Although some S-rich knots show the \hei\ 1.083 \micron\ line,
their intensities are significantly less than those of the He-rich knots.
The Fe-rich knots have strong lines of \feii\ and
some of them have a weak \hei\ 1.083~\micron\ line as well. 
The brightest Fe-rich knot, Knot 10 in Slit 5, also emits 
\feiii\ lines in the {\it K} band.
As in the He-rich knots, 
no lines of C, Si, and S are detected in the Fe-rich knots.

The three knot groups are easily distinguishable
when the flux ratios of \sii\ 1.03~\micron\ multiplets,
\hei\ 1.083~\micron, and \pii\ 1.188~\micron\ are compared
(Figure~\ref{fig-fluxratio}).
The He-rich knots are well separated from the other groups
in F(\hei-1.083)/F(\feii-1.644),
i.e., the ratio is greater than two for the He-rich knots
but lower than two for the S-rich and Fe-rich knots.
The F(\sii-1.03)/F(\feii-1.644) and F(\pii-1.188)/F(\feii-1.644)
of the He-rich knots are mostly smaller than
a few times 1.0 and 0.1, respectively.
Although the S-rich and Fe-rich knots are not clearly distinguished in
the F(\sii-1.03)/F(\feii-1.644) and F(\pii-1.188)/F(\feii-1.644) comparisons,
we see that these ratios are higher in the S-rich knots
than the Fe-rich knots, i.e., F(\sii-1.03)/F(\feii-1.644) $\gtrsim 5$ and
F(\pii-1.188)/F(\feii-1.644) $\gtrsim 0.3$ for the S-rich knots and vice versa.
It is worth noting that
the flux ratios of the knots in Figure~\ref{fig-fluxratio}
and the criteria mentioned above are based on the assumption that
the intrinsic flux ratio of \feii\ 1.257 and 1.644~\micron\ is 1.36
(Section~\ref{sec-pca-method}).
If we adopt different line ratios, e.g., 0.98 to 1.49
\citep[][and references therein]{gia15,koo15},
then the criteria of those three flux ratios will be
F(\hei-1.083)/F(\feii-1.644) = 1.0--2.4, 
F(\sii-1.03)/F(\feii-1.644) = 2.3--6.0, and
F(\pii-1.188)/F(\feii-1.644) = 0.19--0.34.

\subsubsection{Physical Properties of Three Knot Groups}
\label{sec-pca-result-prop}
Figure~\ref{fig-hist} compares the distributions of the knot sizes,
radial velocities, and line widths of the three knot groups.
The angular sizes of most of the knots are in the range 2\arcsec--7\arcsec\
(or 0.03--0.1 pc at a distance of 3.4 kpc)
and there is no significant difference among the three groups in their sizes,
although some of the S-rich and Fe-rich knots are as large as 10\arcsec.
On the other hand, there are significant differences in
their radial velocities and line widths.
The radial speeds of He-rich knots are $\lesssim 200~\kms$,
while those of S-rich knots range from $-2000~\kms$ to $+2000~\kms$
with a median of $+630~\kms$.
The radial velocities of Fe-rich knots range from $-500~\kms$ to $+1500~\kms$
with a median of $+330~\kms$.
In line width,
the He-rich knots have widths of 5--10 \AA,
while the S- and Fe-rich knots have widths of 10--35 \AA.
(Note that our spectral resolution at 1.64 \micron\ is $\sim 6$ \AA.)
Figure~\ref{fig-hist} also compares the distribution of
the \feii\ 1.644~\micron\ line fluxes
among the three knot groups.
While the S-rich and Fe-rich knots have a similar distribution with
an increased number of knots that have faint \feii\ emission,
the pattern is absent in the He-rich knots.
There is no apparent correlation among the
four physical parameters of the knots.

One of the physical parameters of the knots
that can be straightforwardly obtained is electron density
using \feii\ lines originating from levels with similar excitation energies,
because their ratios are mainly determined by electron densities
\citep[e.g.,][]{koo16}.
\feii\ 1.644~\micron\ and 1.677~\micron\ are such lines,
and Figure~\ref{fig-ne} compares their expected ratios as a function of density
for the assumed temperatures of 5000, 10{,}000, and 20{,}000 K (left panel) with
observed values as a function of the \feii\ 1.644~\micron\ flux (right panel).
We see that the ratio is quite insensitive to temperature and
can be used to estimate electron density in the range
$\sim 10^3$--$10^5~{\rm cm^{-3}}$.
The electron density of the He-rich knots is a few $10^4~{\rm cm^{-3}}$,
while the S-rich knots show electron densities over a broad range of 
$10^3~{\rm cm^{-3}}$ to $10^5~{\rm cm^{-3}}$.
The Fe-rich knots have somewhat lower ($10^3$--$10^4~{\rm cm^{-3}}$)
electron densities compared to the other two groups.
There appears to be no correlation between the electron densities and
\feii\ 1.644~\micron\ line fluxes.

\section{Discussion} \label{sec-dis}
In this section, we discuss the origin of the knots
using their spectral characteristics described in the previous section.

\subsection{He-rich and S-rich knots} \label{sec-dis-hek}
The He-rich and S-rich knots have quite distinct spectral properties;
the He-rich knots have high F(\hei-1.083)/F(\feii-1.644)
and low F(\sii-1.03)/F(\feii-1.644), F(\pii-1.188)/F(\feii-1.644),
while the S-rich knots have low F(\hei-1.083)/F(\feii-1.644)
and high F(\sii-1.03)/F(\feii-1.644), F(\pii-1.188)/F(\feii-1.644).
Their kinematic properties are also quite different;
the He-rich knots have low ($\lesssim 200~\kms$) line-of-sight speeds,
while the S-rich knots have high (up to $\sim 2000~\kms$) line-of-sight speeds.
These spectral and kinematical properties suggest that 
the He-rich knots are dense, slow-moving CSM
swept up by the SN blast wave,
while the S-rich knots are fast-moving SN ejecta that have been shocked.
The same conclusion was reached by \citet{koo13},
who performed an abundance analysis
using \pii\ 1.188~\micron\ and \feii\ 1.257~\micron\ lines.
They showed that the relative abundance of P
(a major product of the stellar Ne-burning layer)
to Fe (in number) for the He-rich knots is similar to the solar abundance,
whereas that of the S-rich knots is 10--100 times higher than
the solar abundance.

The characteristics of the two types of knots match 
well with those of QSFs and FMKs known from previous optical studies
(see Section~\ref{sec-int} for a summary of their properties).
Similar to the optical QSFs,
the He-rich knots have bright \hei\ lines
together with \hi\ and \nn\ lines;
all the He-rich knots have a \hei\ 1.083 \micron\ line,
6 out of the 7 show \hi\ Pa$\beta$, Br$\gamma$ lines,
and the three brightest ones also have \nn\ 1.040, 1.041 \micron\ lines.
Like the optical FMKs dominated by lines of ionized heavy elements O, S, Ar,
the S-rich knots show strong S lines plus \pii\ and \silvi\ lines;
all the S-rich knots have \sii\ 1.03 \micron\ multiplets,
and 39 out of 45 S-rich knots also have
\pii\ 1.188 \micron\ and/or \silvi\ 1.963 \micron\ lines.
Indeed, the overall spectra of He-rich and S-rich knots are
similar to the NIR spectra of QSFs and FMKs
obtained by \citet{ger01}, respectively.
Furthermore, the radial velocities of the two NIR knot groups
are well consistent with those of the optical groups.
For example, the radial velocity of the He-rich knots is $-50 \pm 90~\kms$,
while the generally accepted radial velocity of QSFs is 
$-140 \pm 300~\kms$ \citep{van85,ree95}.
In addition,
the median line-of-sight velocity of S-rich knots is $+620~\kms$,
while the systematic velocity of the FMKs
is $+770\pm40~\kms$ \citep{ree95}.

Rich emission lines of Si, P, and S
in the S-rich knots and FMKs
imply that they are the SN ejecta originated from the Ne- and O-burning layers.
The two bright emission lines, \pii\ 1.188~\micron\ and \feii\ 1.644~\micron,
have comparable excitation energies and critical densities,
and so their line ratios are strongly dependent on their abundance ratio,
$X$(P/Fe) \citep{koo13}.
As seen in Figure~\ref{fig-fluxratio}, 
there is a large scatter in this line ratio for S-rich knots,
which implies that the abundance ratio $X$(P/Fe) varies
almost two orders of magnitude for these knots.
We also found that 13 out of 45 S-rich knots have clear
but relatively weak emission lines of \hei\ and/or \ci.
The detection of the He, C, and Fe lines in the S-rich ejecta,
which are either lighter or heavier elements than the O-burning materials,
might infer microscopic mixing during the SN explosion.
In many S-rich knots,
a highly ionized Si line, \silvi\ 1.963 \micron, is detected,
while in a few S-rich knots, a \sili\ 1.645 \micron\ line is also detected.
The detection of Si in very different ionization stages
indicates the broad range of temperatures in the S-rich knots.

\subsection{Fe-rich knots} \label{sec-dis-fek}
In Section~\ref{sec-pca-result},
we found that the Fe-rich knots exhibit intermediate characteristics between
He-rich and S-rich knots; they emit strong \feii\ lines
without any Si, P, and S lines,
but have high line-of-sight speeds of up to $\sim 1500~\kms$. 
A few knots also emit an \hei\ 1.083~\micron\ line.
The high velocities, however, suggest that
they are not dense QSFs represented by the He-rich knots.
Their line widths are also considerably broader than those of He-rich knots,
i.e., 10--35~\AA\ vs. 5--10~\AA\ (Figure~\ref{fig-hist}).
On the other hand, the missing Si, P, and S lines indicate that
the abundances of these Ne- and O-burning elements are
very low in these Fe-rich knots.
We can consider two possible explanations
regarding the origin of the Fe-rich knots:
(1) swept-up CSM around contact discontinuity (CD) or
(2) shocked SN ejecta enriched with Fe elements
that had been synthesized in explosive Si burning.
The ambient medium that the Cas A SN blast wave is propagating into 
is believed to be CSM with an $r^{-2}$
density distribution \citep[e.g.,][]{leejj14}.
In such a case,
1D similarity solutions show that
the shocked CSM accumulates at the CD with infinite density asymptotically
\citep{che82}.
We thus expect ``dense'' CSM expanding at a speed comparable to
the shocked SN material.
In the real situation, however,
this interacting region between the shocked SN ejecta and the shocked CSM is
hydrodynamically unstable and distorted,
with the density of the shocked CSM limited to $\lesssim 10$ times the density
at the ambient shock \citep{che92,blo01,van09}.
The temperature of the shocked CSM near the CD, therefore,
may be lower than the typical temperature ($\sim 2$ keV) of the shocked CSM
\citep{hwa12},
but probably no more than a factor of 10,
and all of the Fe in the shocked CSM will be in high ionization stages.
Furthermore, \hi\ lines are not detected in all Fe-rich knots
with an upper limit of F(\hi-Pa$\beta$)/F(\feii-1.257) $\lesssim 0.1$.
Note that the observed ratio of these line intensities
ranges between 0.05 and 10 for SNRs
while it is $\sim 50$ for Orion,
which should be the typical ratios for shocked and photoionized gases
of cosmic abundance, respectively \citep{koo15}.
Therefore, H must be depleted in Fe-rich knots.
The non-detection of He and N lines might also indicate that
the abundances of these `circumstellar' elements are very low
in Fe-rich knots,
although this needs to be confirmed from other waveband observations.
(See the next paragraph for an explanation of the faint He lines
detected in some Fe-rich knots.)
We may therefore conclude that the Fe-rich knots are not likely the shocked CSM.

This leaves the second possibility,
i.e., the Fe-rich knots are Fe-enriched SN ejecta.
The high velocities and large velocity widths are consistent with 
SN ejecta being swept up by the reverse shock. 
The low \pii\ and \sii\ fluxes compared to the \feii\ flux,  
however, implies that the abundances of P and S, which are  
Ne- and O-burning materials, are very low,
which is in sharp contrast to the S-rich knots.
These characteristics strongly suggest that 
{\em the Fe-rich knots are most likely ``pure'' Fe ejecta
synthesized in the deepest stellar interior.}
The weak \hei\ 1.083 \micron\ line detected in some Fe-rich knots
could be due to an $\alpha$-rich freeze-out process during the
explosive Si burning;
just after the explosion,
complete Si burning with an $\alpha$-rich freeze-out
occurs under high temperatures and low density conditions
in the stellar deep layer,
and many alpha particles are ``frozen out''
without participating in further nucleosynthetic processes
\citep{woo73}.
Similar dense, Fe-predominant ejecta,
likely from the $\alpha$-rich freeze-out process,
have been detected in another young core-collapse SNR G11.2-0.3
\citep{moo09}.

Pure Fe ejecta have been detected in X-rays (see below)
but not in optical or NIR wavebands.
This is surprising considering the extensive optical/NIR studies
carried out for Cas A since its discovery. 
Figure~\ref{fig-knotpos} partly gives an answer.
In the right panel of Figure~\ref{fig-knotpos},
red is an \feii\ 1.644 \micron\ narrow-band image
while green and blue are {\it Hubble Space Telescope} ({\it HST})
ACS/WFC F850LP and F775W images 
which are dominated by ionized S and O lines, respectively
\citep{fes06a,ham08}.
Previous optical observations had been mostly toward 
the northern ejecta shell bright in 
ionized O, S, and Ar lines \citep[e.g.,][]{che79,hur96,fes01b} or 
toward FMKs outside the main shell \citep[e.g.,][]{fes88,fes06a,fes01a}.
Figure~\ref{fig-knotpos}, however, shows that
the southwestern (SW) main ejecta shell, which is bright in the \feii\ line
but faint in the ionized O and S lines,
is the region where Fe-rich ejecta can be found.
Indeed, our result shown in the left panel of Figure~\ref{fig-knotpos}
confirms this;
9 out of 11 Fe-rich knots are located in Slits 4--6.
(Note that the compact red knots in the interior and beyond the SW shell
are mostly QSFs,
and they are bright in \nii\ 6548, 6583 \AA\ and H$\alpha$ images too
\citep{van85,ala14}.)
The slit positions are determined from an \feii\ 1.644 \micron\ image,
so that some of them were placed
toward the red portions of the main ejecta shell and,
by decomposing the emission into individual velocity components,
we were able to identify Fe-rich ejecta.
It is worth noting that
\citet{rho03} also noted the bright \feii\ emission in the SW shell
in their \feii\ 1.644 \micron\ image of Cas A.
Meanwhile, we can see that
the \feii\ 17.9 \micron\ emission is much brighter than the emission from
O-burning elements such as Ar and S in the SW shell
in the {\it Spitzer} mid-infrared maps of ionic lines \citep{enn06,smi09}.
Our result suggests that this \feii\ emission-predominant area,
i.e., the red area of the SW ejecta shell in Figure~\ref{fig-knotpos},
is probably mainly composed of Fe ejecta.

It is not easy to identify the counterpart of Fe-rich knots in optical
images because several velocity components are usually superposed
along the line of sight toward the main ejecta shell.
The brightest Fe-rich knot (Knot 10 in Slit 5; hereafter K10), however,
is somewhat isolated and we can identify its counterpart.
In Figure \ref{fig-k10},
the upper two images are \feii\ 1.644 \micron\ images at different epochs
and they show that K10 is a clump of $\sim 10\arcsec \times 3\arcsec$
elongated along the slit.
The two \feii\ images clearly show that
the clump is moving fast tangentially.
The proper motion is measured $0\farcs28$ yr$^{-1}$,
implying a tangential velocity of $4500 \pm 200~\kms$.
In the lower F775W and F850LP images,
we see diffuse faint emission at the position of K10 (see the red contour).
Its brightness distribution is different,
with two small ($\lesssim 2\arcsec$) bright spots (S1 and S2 in the figure)
in the lower part of the clump.
One of these bright spots, S1, is coincident with an S-rich knot
(Knot 9 in Slit 5),
which is spatially coincident with K10
but has a line-of-sight velocity ($+1500~\kms$)
that is very different from that of K10 ($-300~\kms$).
The other bright spot, S2, must also be due to an S-rich knot
not detected in our spectroscopy.
So excluding these two bright knots,
K10 appears faint in F775W and F850LP images.
We suspect that most of the diffuse emission in the F775W and F850LP images
is due to optical \feii\ lines.
This can be confirmed by optical spectroscopy.
Recently, there have been optical spectral mapping observations of Cas A 
\citep{ree95,mil13,ala14} and, in principle,    
a similar analysis can be done to detect Fe-rich knots,
although the optical \feii\ lines,
e.g., \feii\ 7155 and 8617 \AA\ lines, will be much fainter than
the \feii\ 1.644 \micron\ line
because of large interstellar extinction ($A_{\rm V} = 5$--10 mag) toward Cas A
\citep[e.g.,][]{eri09,hwa12,leeyh15}.

The distribution and amount of Fe ejecta in Cas A
have been a subject of controversy.
Previous X-ray studies detected
hot and diffuse ``pure'' Fe ejecta with mass $\sim 0.1~\msun$
that might have formed by $\alpha$-rich freeze-out
during the complete Si burning \citep{hwa03,hwa12}.
These X-ray-emitting shocked Fe ejecta are distributed mainly
in the southeastern and northern regions of the remnant
(Figure~\ref{fig-fedist}).
On the other hand,
the hard X-ray emission from the radioactive decay of $^{44}$Ti
has been detected in the interior of the main ejecta shell
\citep[Figure~\ref{fig-fedist};][]{gre14}.
Since $^{44}$Ti is essentially synthesized in complete Si burning with
$\alpha$-rich freeze-out in the innermost region \citep[e.g.,][]{mag10},
$^{44}$Ti traces ``pure'' $^{56}$Ni or its stable nuclei $^{56}$Fe.
The majority of the observed $^{44}$Ti is inside the reverse shock and
therefore from unshocked Fe ejecta.
The inferred mass of the unshocked Fe ejecta is
$\sim 0.1~\msun$ \citep{gre14}.
Such unshocked Fe ejecta, however, have not yet been detected,
presumably because they are cool
\citep[$T \lesssim 40$ K; e.g.,][]{bar10,sib10,leeyh15}.
Instead, the X-ray-emitting Fe ejecta
located just outside the $^{44}$Ti emission was attributed to
the corresponding shocked ejecta \citep[Figure~\ref{fig-fedist};][]{gre14}.
However, the missing X-ray-emitting Fe ejecta
toward the south and northeast directions from the explosion center
have been puzzling.
We note that the NIR \feii-bright, red portion of the SW shell
appears to be in contact with the interior $^{44}$Ti-emitting region
(Figure~\ref{fig-fedist}).
This might be the case
for the small red portion near Slit 2 in the northeastern shell, too.
Therefore, if these \feii-bright regions are composed of
shocked, dense Fe ejecta, as implied from our spectroscopic result,
it explains why we do not see X-ray-emitting diffuse Fe ejecta
toward these directions;
the unshocked ``pure'' Fe ejecta traced by
the radioactive $^{44}$Ti emission in the interior of the main shell
is composed of both dense and diffuse ejecta,
and when these are swept up by a reverse shock,
we observe either NIR \feii\ emission or X-ray emission
depending on their densities.
We do not see NIR emission associated with
the central bright $^{44}$Ti emission
but this could be because the shock is face-on, as suggested by
the large red-shifted central velocity (1100--3000 $\kms$)
of the $^{44}$Ti line \citep{gre14}, or
because the reverse shock has not yet reached the dense, unshocked Fe ejecta.
Recent multi-dimensional simulations show that
such global asymmetry in Fe (or $^{56}$Ni) density can arise from
the low-mode convection of the innermost region just after the core bounce
\citep[e.g.,][]{won13}.
Future NIR spectral mapping observations
revealing the 3D distribution of the dense Fe ejecta
will be helpful for understanding the SN explosion dynamics of
the innermost region.

\section{Summary} \label{sec-sum}
We have carried out NIR spectroscopic observations toward
the main ejecta shell of the young SNR Cas A.
In total, 63 individual knots were identified from eight slit positions
by using a clump-finding algorithm.
Each of these knots has distinct kinematical and spectral properties.
Within the {\it JHK} spectral range (0.94--2.46 \micron),
we found 46 emission line features including a dozen bright \feii\ lines,
forbidden lines of other metallic species, and H and He lines.
We employed the PCA method to classify the knots
based on their relative line fluxes into three distinctive groups:
He-rich knots of pre-supernova circumstellar wind material,
plus S-rich and Fe-rich knots of SN ejecta material.
The He-rich and S-rich knots correspond to QSFs and FMKs
studied in the visible waveband,
while Fe-rich knots, showing in general only \feii\ emission lines,
are likely `pure' dense Fe ejecta from the innermost layer of the progenitor.
We summarize our main results as follows.

\begin{figure}
\center{
\includegraphics[scale=0.9]{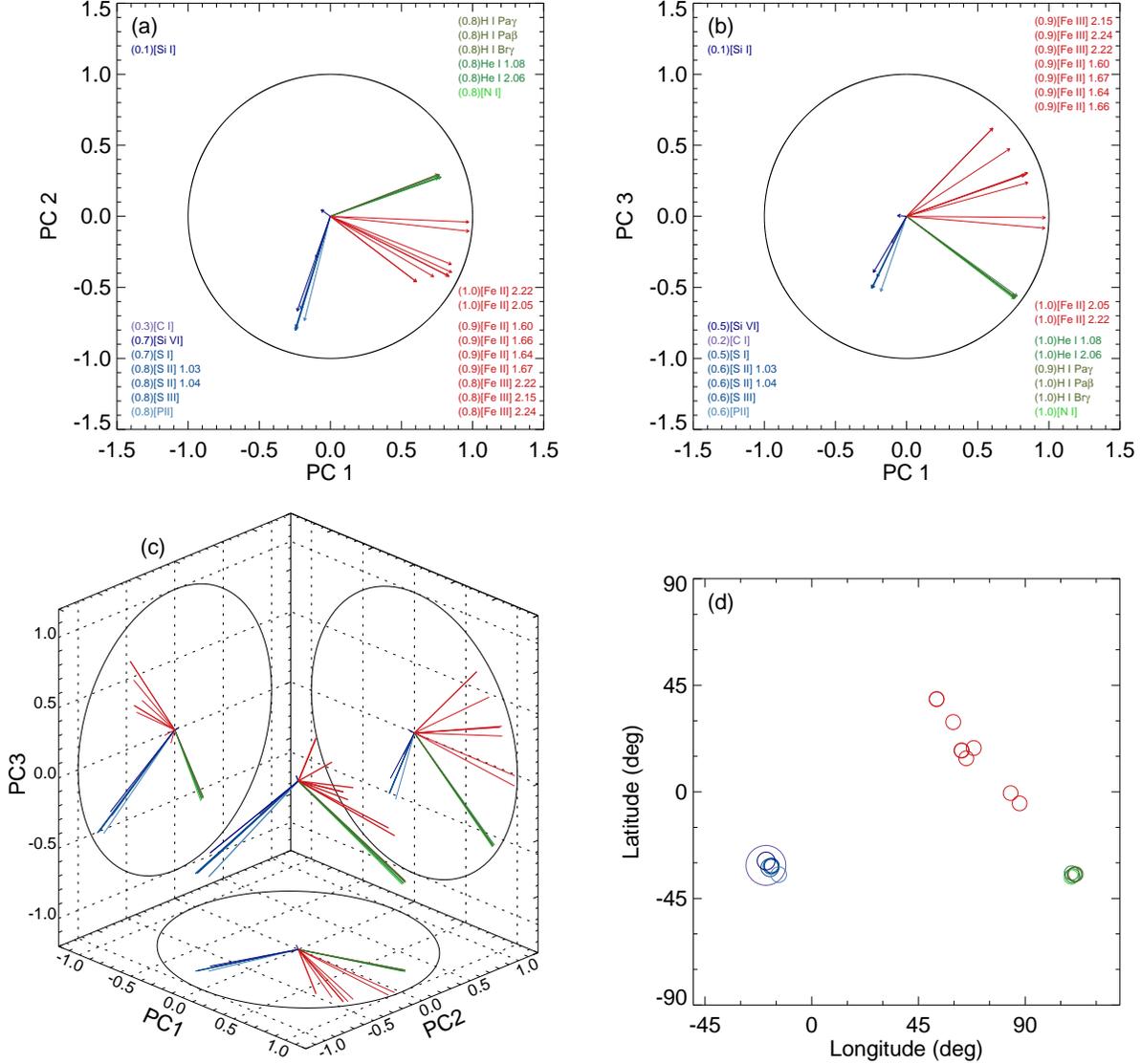}
}
\caption{
	Projections of the 23 attributes (or spectral lines)
	on the plane of	(a) PC1-PC2 and (b) PC1-PC3.
	A circle of unity radius is overplotted for comparison.
	The number within parentheses at the corners of (a) and (b)
	is radius from the center,
	that represents the normalized contribution of the line
	at each PC plane.
	The 3D projections and the projections within a sphere are
	in (c) and (d), respectively (see the text for details).
	The reference position in (d),
	where (Longitude, Latitude) = (0\degr, 0\degr),
	corresponds to (PC1, PC2, PC3) = (0, -1, 0),
	and Longitude and Latitude are measured toward east and north,
	respectively.
	The size of circles in (d) is inversely proportional to
	the length of the vector in (c).
	(An interactive 3D pdf version and it source code
	written in {\sc Asymptote}
	are available in a tar.gz package in the electronic journal.)
} \label{fig-atr}
\end{figure}

\begin{figure}
\center{
\includegraphics[scale=0.9]{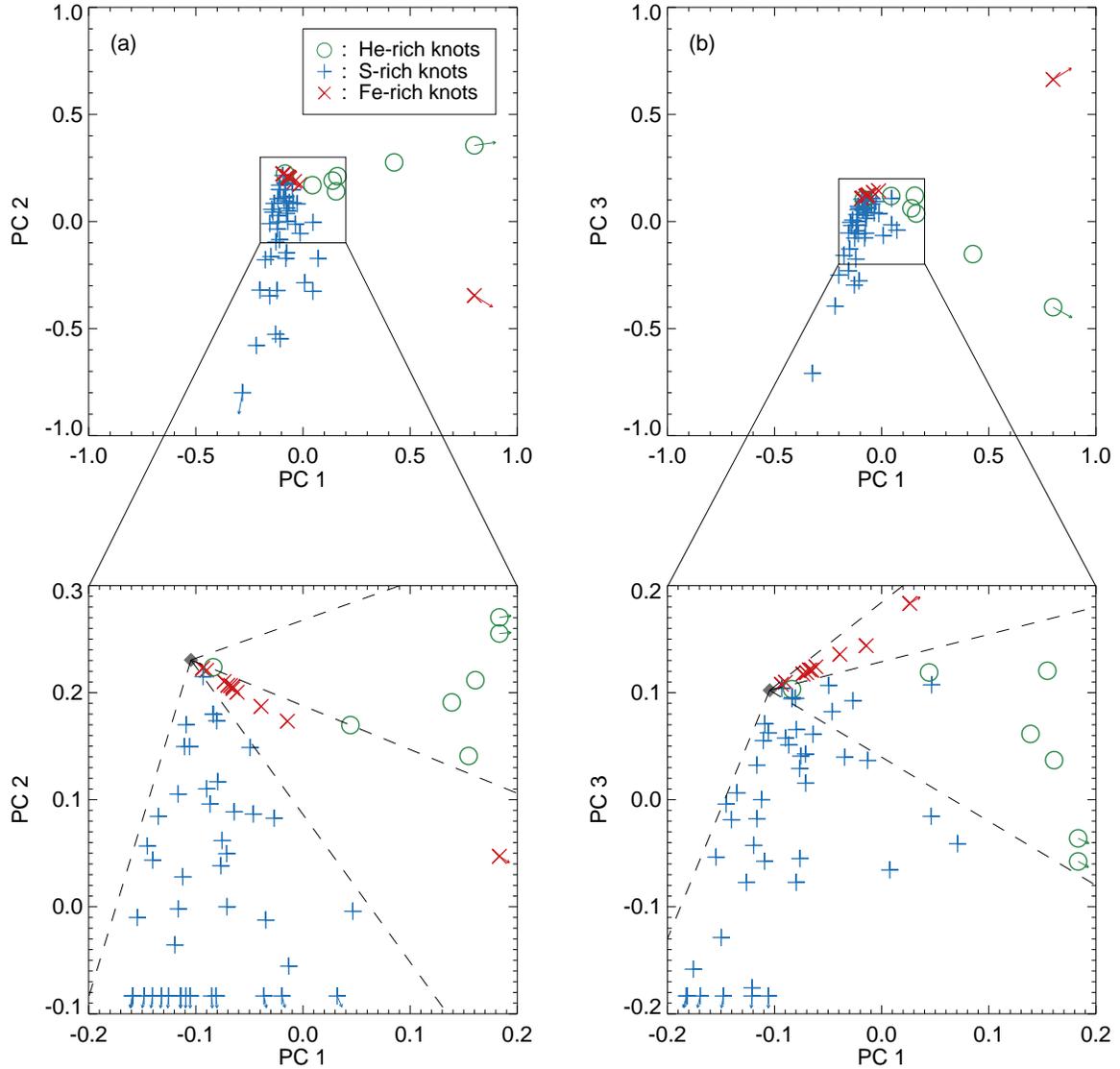}
}
\caption{
	Projections of the 63 objects (or knots) on the plane of
	(a) PC1-PC2 and (b) PC1-PC3.
	The green open circles, blue crosses, and red X symbols
	indicate He-rich, S-rich, and Fe-rich knots, respectively.
	The enlarged views of the crowded central regions
	are shown in the lower panels.
	The black dashed lines in the lower panels represent
	the criteria of the groups,
	and the black diamond indicates the convergent point
	(see text for details).
	(An interactive 3D pdf version and it source code
	written in {\sc Asymptote}
	are available in a tar.gz package in the electronic journal.)
} \label{fig-obj}
\end{figure}

\begin{figure}
\center{
\includegraphics[scale=0.9]{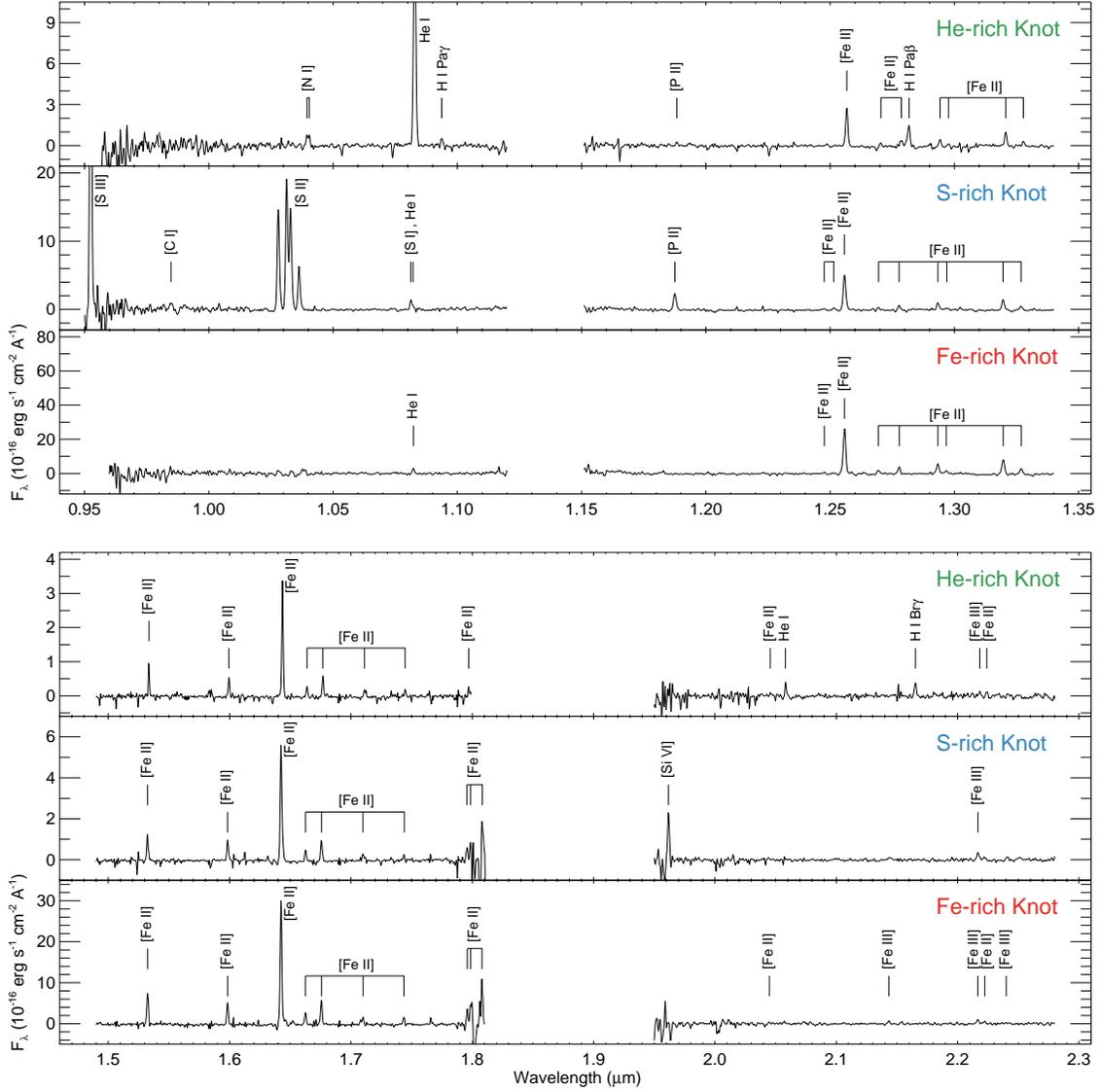}
}
\caption{
	Sample 1D spectra of He-rich (Knot 5 in Slit 7),
	S-rich (Knot 3 in Slit 3),
	and Fe-rich (Knot 10 in Slit 5) knots.
	The spectra have been smoothed by a Gaussian kernel
	with an FWHM of 1 pixel (2--3 \AA).
} \label{fig-1dspec}
\end{figure}

\noindent
1. The PCA showed that the NIR spectral lines
can be grouped into three groups:
(1) Group 1, composed of \hi\ and \hei\ lines
together with \nn\ lines,
(2) Group 2, composed of forbidden lines of Si, P, and S, and 
(3) Group 3, composed of forbidden Fe lines.
The lines in the first two PCs are strongly correlated with each other,
while the correlation is rather weak among the forbidden Fe lines in Group 3.
These three spectral groups of the emission lines are almost independent in
3D PC space (Figure~\ref{fig-atr}).

\begin{figure}
\center{
\includegraphics[scale=0.9]{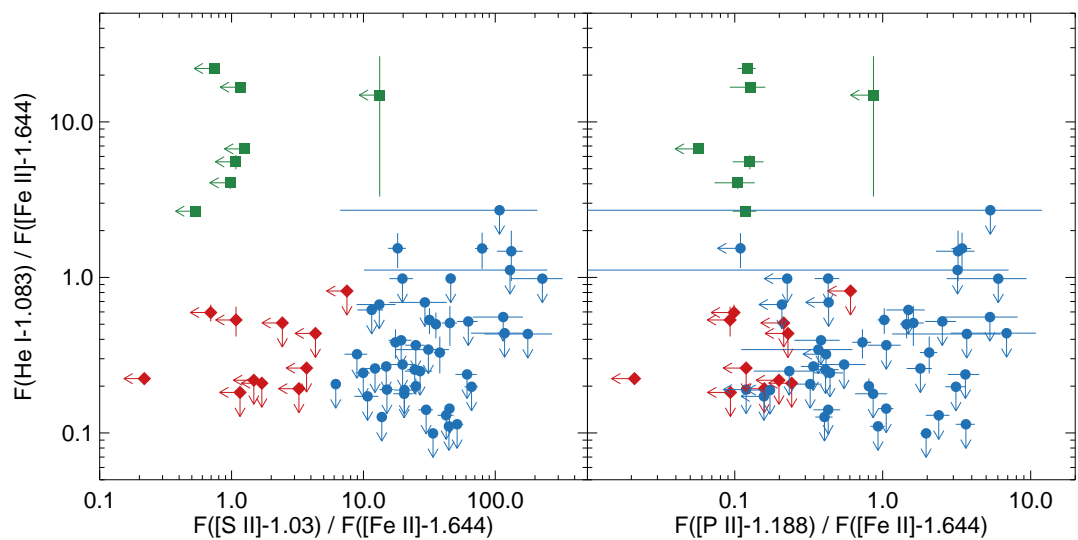}
}
\caption{
	Flux comparisons among the lines of \hei\ (1.083~\micron),
	\sii\ (1.03~\micron\ multiplets), and \pii\ (1.188~\micron).
	The extinctions are corrected
	and the fluxes are normalized by
	the \feii\ 1.644~\micron\ line fluxes.
	The three knot groups are represented by
	green squares (He-rich knots),
	blue circles (S-rich knots), and red diamonds (Fe-rich knots).
	The arrows represent $3\sigma$ upper limits.
} \label{fig-fluxratio}
\end{figure}

\begin{figure}
\center{
\includegraphics[scale=0.9]{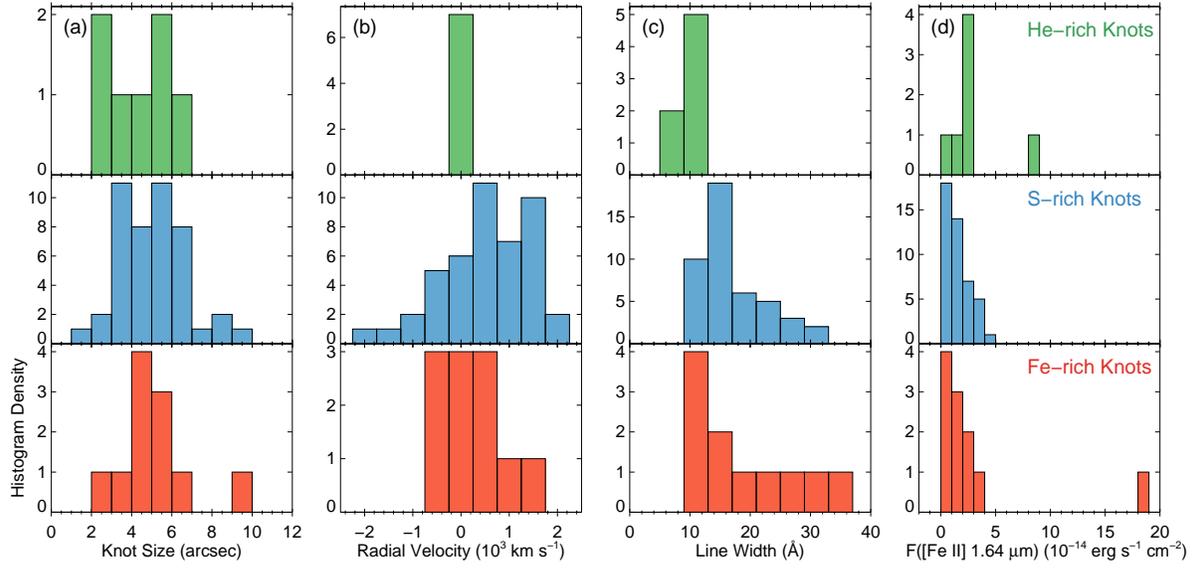}
}
\caption{
	Histograms of (a) knot size, (b) radial velocity, (c) line width,
	and (d) extinction-corrected \feii\ 1.644~\micron\ flux.
} \label{fig-hist}
\end{figure}

\begin{figure}
\center{
\includegraphics[scale=0.9]{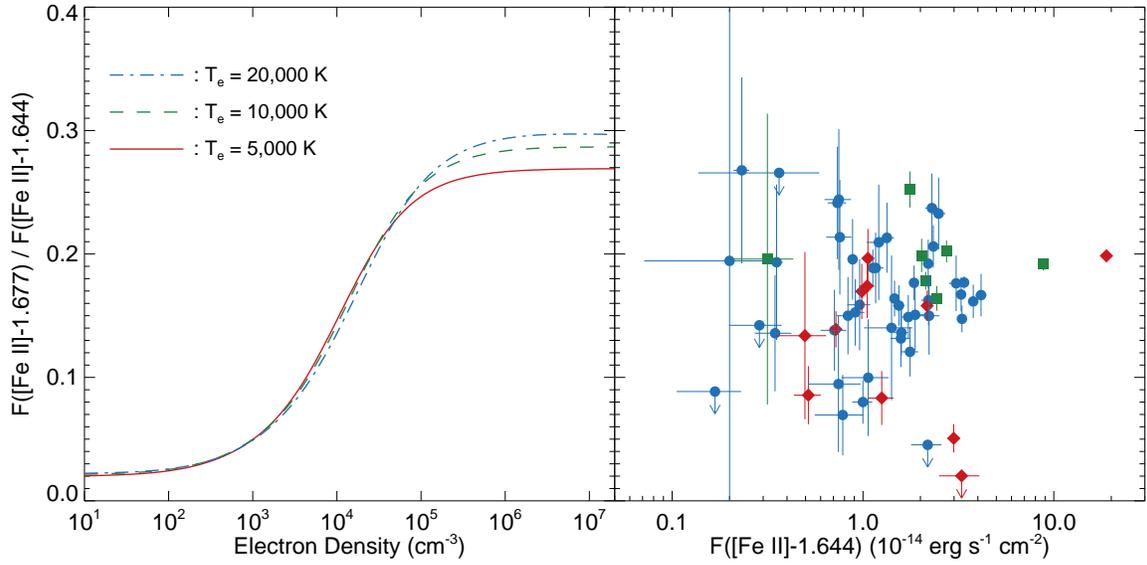}
}
\caption{
	(Left) F(\feii-1.677)/F(\feii-1.644) ratio
	as a function of electron density for gas
	in statistical equilibrium at
	$T_e=$ 5000 K, 10{,}000 K, and 20{,}000 K.
	(Right) F(\feii-1.677)/F(\feii-1.644) vs.
	F(\feii-1.644) of the knots.
	The symbols are the same as in Figure~\ref{fig-fluxratio}.
	All fluxes are extinction-corrected.
} \label{fig-ne}
\end{figure}

\begin{figure}
\center{
\includegraphics[scale=0.9]{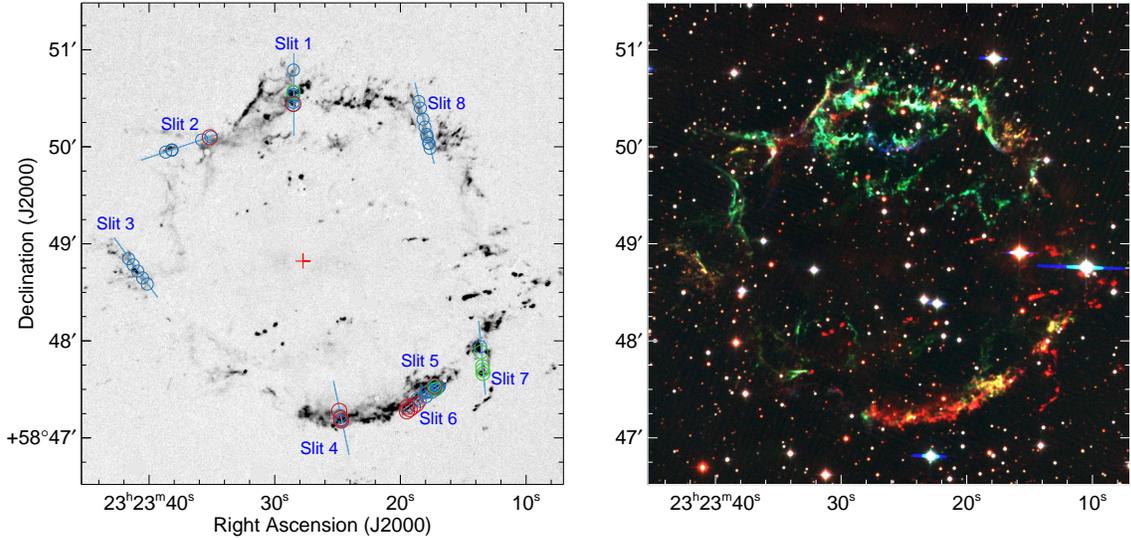}
}
\caption{
	(Left) Locations of 63 knots in Table~\ref{tbl-knot}.
	(See Figure~\ref{fig-clumps} for the knot numbers in each slit.)
	The He-rich, S-rich, and Fe-rich knots are marked in
	green, blue, and red, respectively.
	The background is the continuum-subtracted
	\feii\ 1.644 \micron\ narrow-band image in Figure~\ref{fig-slit}.
	(Right) A three-color composite image of Cas A with
	an \feii\ 1.644 \micron\ narrow-band image in red,
	and {\it HST} ACS/WFC F850LP and F775W images in green and blue.
	The {\it HST} images are dominated by
	\siii\ 9069, 9531 \AA\ and \sii\ 1.03 \micron\ multiplets (F850LP)
	and \oii\ 7319, 7330 \AA\ lines  (F775W), respectively
	\citep{fes06a,ham08}.
	The \feii\ 1.644 \micron\ image was observed in 2005 August.
	The {\it HST} ACS/WFC images are from the Hubble Legacy Archive
	(https://hla.stsci.edu)
	and they were taken in 2004 December \citep{fes06a}.
	To match their angular resolutions,
	the {\it HST} images were smoothed by using a Gaussian kernel
	with a FWHM of 0\farcs9.
} \label{fig-knotpos}
\end{figure}

\begin{figure}
\center{
\includegraphics[scale=0.9]{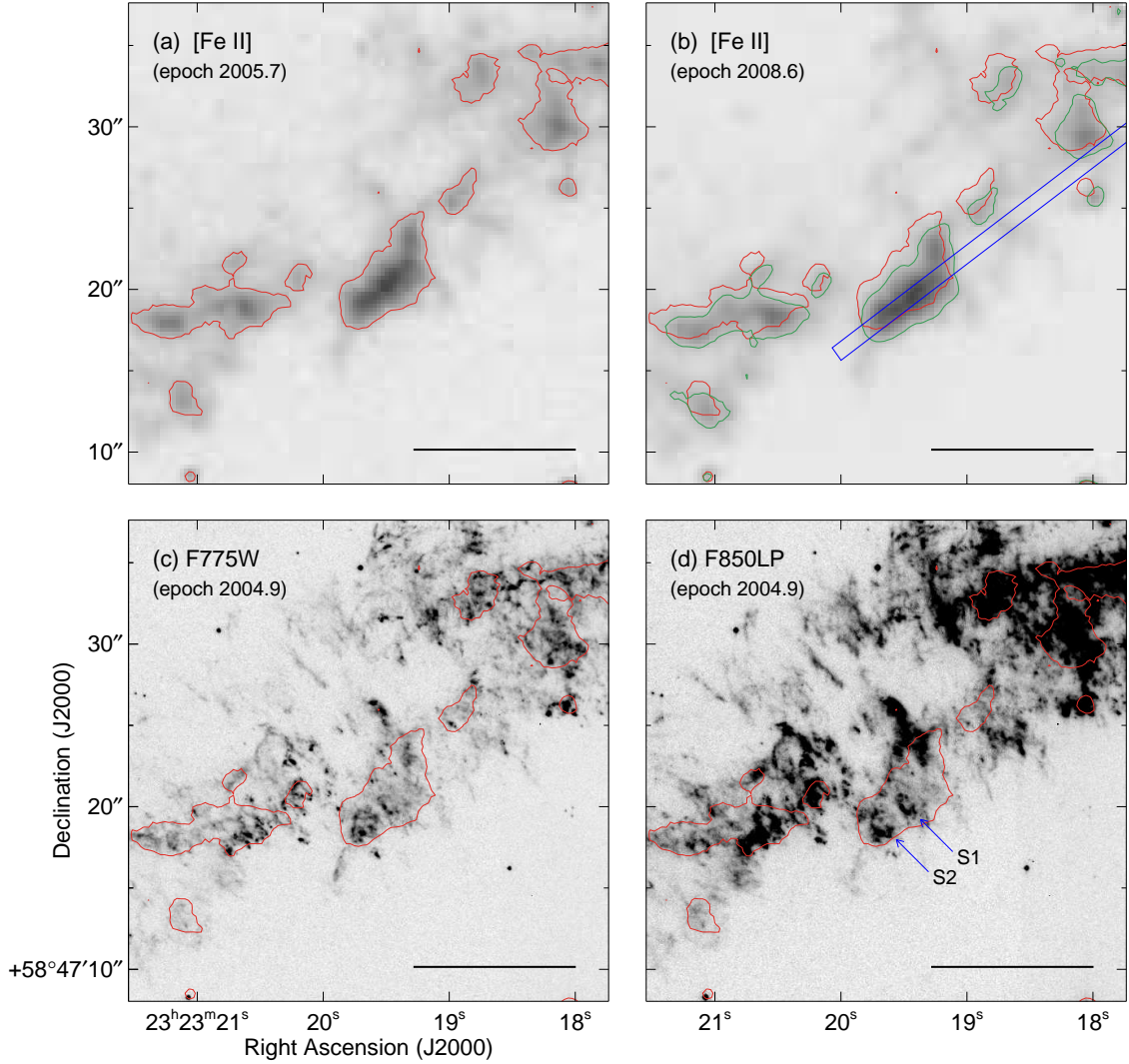}
}
\caption{
	Enlarged views of the SW shell where the brightest Fe-rich knot
	(Knot 10 in Slit 5; K10) has been detected.
	(a)--(b) \feii\ 1.644 \micron\ narrow-band images obtained in
	2005 August and in 2008 August.
	The red and green contours indicate the brightness level of
	$1 \times 10^{-4}~\ergscmsr$ in the 2005 and 2008 images, respectively.
	The blue bar in the 2008 image represents the position of Slit 5
	across Knot 10.
	(c)--(d) {\it HST} ACS/WFC F775W and F850LP images taken in 2004 December.
	The red contours overlaid on the images are the same as in (a).
	(d) S1 and S2 are the two bright knots
	mentioned in Section~\ref{sec-dis-fek}.
	The angular resolutions are 0\farcs1 and 0\farcs9 for
	the optical and \feii\ images, respectively.
	The black scale bar in the lower right of each panel represents
	an angular scale of 10\arcsec,
	and the low and high thresholds of the grayscale are
	$-5 \times 10^{-5}$ and $5 \times 10^{-4}~\ergscmsr$ in all images.
} \label{fig-k10}
\end{figure}

\begin{figure}
\center{
\includegraphics[scale=0.9]{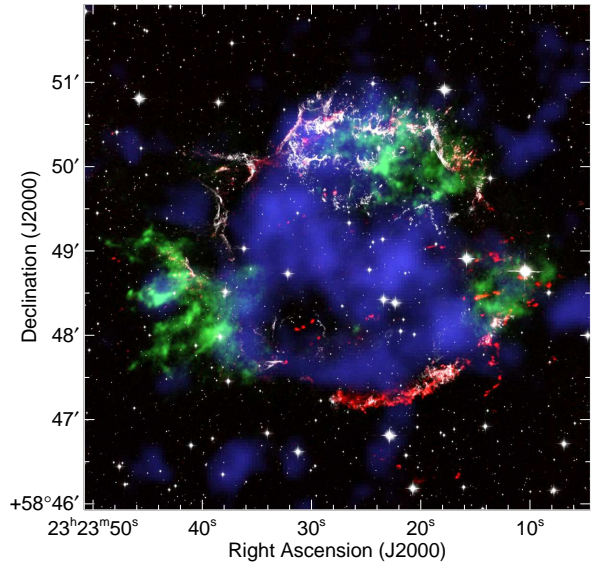}
}
\caption{
	A four-color composite image of Cas A with
	the \feii\ 1.644 \micron\ narrow-band image in Figure 1 in red,
	the {\it Chandra} Fe K-shell (6.52--6.94 keV) image in green
	\citep{hwa04},
	the {\it NuSTAR} hard X-ray $^{44}$Ti (67.9 and 78.4 keV) image in blue
	\citep{gre14},
	and the {\it HST} ACS/WFC F850LP image in white \citep{fes06a}.
} \label{fig-fedist}
\end{figure}

\noindent
2. The distribution of the knots in the PC planes matches well with 
the above spectral groups,
and we classified the knots into three groups:
(1) He-rich, (2) S-rich, and (3) Fe-rich knots. 
The knots belonging to these three groups are well separated
from each other in
F(\sii-1.03)/F(\feii-1.644) vs. F(\hei-1.083)/F(\feii-1.644) plane
(Figure~\ref{fig-fluxratio}),
so that one may use these line ratios to classify the knots in Cas A.
It would be interesting to determine
whether this classification methodology applies for other core-collapse SNRs.

\noindent
3. The He-rich knots show 
bright emission lines of \hei\ 1.083~\micron\ and \feii\
together with \nn\ and \hi\ lines.
Their line-of-sight speeds are small ($\lesssim 200~\kms$).
From these chemical and kinematical characteristics,
we conclude that the He-rich knots are dense CSM
swept up by the SN blast wave.
These knots correspond to the previously known QSFs.

\noindent
4. The S-rich knots show strong forbidden lines of S
together with \pii\ and \silvi,
and their line-of-sight speeds reach a few 1000 $\kms$.
These chemical and kinematical properties indicate that
the S-rich knots are dense SN ejecta material
mostly originating from the O-burning layers
and swept up by a reverse shock.
These knots correspond to the FMKs
detected in previous optical studies.

\noindent
5. The Fe-rich knots only show strong \feii\ and \feiii\ lines,
and no or weak \hei\ 1.083~\micron\ lines.
Like the S-rich knots,
they have large line-of-sight speeds (up to $\sim 1500~\kms$)
and broad line widths (10--35 \AA),
but they do not show the lines from Si, P, and S.
Some Fe-rich knots show \hei\ 1.083~\micron\,
but their fluxes compared to the \feii\ lines are much weaker than
those of the He-rich knots.
These spectroscopic properties suggest that
the Fe-rich knots are most likely ``pure'' dense Fe ejecta
from the innermost layer of the SN.
The comparison of \feii\ 1.644 \micron\ images with
the {\it HST} ACS/WFC F850LP and F775W
and {\it NuSTAR} $^{44}$Ti images reveals that
these Fe ejecta are mainly distributed in the SW main ejecta shell,
just outside the unshocked $^{44}$Ti in the interior.
This supports that
there could be a large amount of unshocked ``pure'' Fe ejecta
associated with $^{44}$Ti.
Together with the diffuse, X-ray-emitting ``pure'' Fe ejecta
detected by {\it Chandra},
our result implies that
the Fe ejecta synthesized in the innermost region develop
large-scale non-uniformity during the SN explosion
and are expelled asymmetrically.
This seems to be consistent with
the low-mode, convection-driven SN explosion model.

\acknowledgments
We wish to thank the anonymous referee
for the very useful comments and suggestions
which helped us improve the quality of the paper.
We thank Brian Grefenstette and Fiona Harrison
for providing the {\it NuSTAR} data.
We also want to thank John Raymond and Sung-Chul Yoon
for helpful discussions.
The interactive 3D figures were made by using {\sc Asymptote}
which is a descriptive vector graphics language\footnote{\url{http://asymptote.sourceforge.net}}.
This research was supported by Basic Science Research Program through
the National Research Foundation of Korea(NRF)
funded by the Ministry of Science, ICT and future Planning
(2014R1A2A2A01002811).

{}

\clearpage

\clearpage

\clearpage

\clearpage

\clearpage

\clearpage

\clearpage

\clearpage

\clearpage

\clearpage

\clearpage

\clearpage

\clearpage
\begin{deluxetable}{cccrccr}
\tablewidth{0pt}
\tablecaption{Parameters of the Spectroscopic Observations \label{tbl-log}}
\tablehead {
\colhead{Date} & \colhead{Slit} 
   & \colhead{Central Coordinate}
   & \colhead{P.A. \tablenotemark{a}} &  \colhead{Mode \tablenotemark{b}}
   & \colhead{Slit Length \tablenotemark{c}} & \colhead{Exposure} \\
\colhead{(UT)} & \colhead{} & \colhead{$\alpha$(J2000) ~~~~~ $\delta$(J2000)}
   & \colhead{(deg)} & \colhead{} & \colhead{(\arcsec)} & \colhead{(s)}}
\startdata
2008 Jun 29 & 1                    & 23:23:28.54  +58:50:32.2 &  0.0  & Both    & 50.50 & 300 $\times$ 4 \\
            & 8                    & 23:23:18.13  +58:50:14.4 & 13.4  & AB      & 51.00 & 300 $\times$ 5 \\
            & 2                    & 23:23:37.60  +58:49:59.7 & 288.7 & AB      & 50.25 & 300 $\times$ 4 \\
            & ST\tablenotemark{d}  & 23:48:53.97  +59:58:44.3 &  0.0  & AB      & 30.00 &  30 $\times$ 6 \\
2008 Aug 08 & 3                    & 23:23:41.11  +58:48:45.2 & 36.1  & AB      & 45.25 & 300 $\times$ 6 \\
            & 4                    & 23:23:24.77  +58:47:12.2 & 12.0  & AB      & 46.25 & 300 $\times$ 6 \\
            & 5                    & 23:23:18.50  +58:47:25.1 & 307.3 & OS      & 30.00 & 300 $\times$ 1 \\
            & 6                    & 23:23:18.33  +58:47:23.5 & 307.3 & OS      & 30.00 & 300 $\times$ 1 \\
            & 7                    & 23:23:13.57  +58:47:49.2 & 5.2   & AB      & 45.25 & 300 $\times$ 6 \\
            & ST\tablenotemark{d}  & 23:48:53.97  +59:58:44.3 &  0.0  & AB      & 30.00 & 30 $\times$ 8 \\
\enddata
\tablenotetext{a}{Position angle (counterclockwise from north on the plane of the sky)}
\tablenotetext{b}{AB: ABBA nodding pattern, OS: Object-Sky pattern, Both: combined AB and OS}
\tablenotetext{c}{Effective Slit Length = instrument slit length (30\arcsec) + Nodding offset length}
\tablenotetext{d}{Standard Star, HD223386, for flux calibration}
\end{deluxetable}

\clearpage
\begin{deluxetable}{cc|c|rcrrr}
\tablewidth{0pt}
\tabletypesize{\tiny}
\tablecaption{Physical Parameters of 63 Identified Knots \label{tbl-knot}}
\tablehead {
\colhead{Slit}
   & \colhead{Knot}
   & \colhead{Central Position}
   & \colhead{Size\tablenotemark{a}}
   & \colhead{Knot}
   & \colhead{$A_{\rm V}$\tablenotemark{c}}
   & \colhead{$v_{\rm rad}$}
   & \colhead{[Fe II] 1.644 \micron\ Flux\tablenotemark{d}} \\
\colhead{No.}
   & \colhead{No.}
   & \colhead{$\alpha$(J2000) ~ $\delta$(J2000)}
   & \colhead{(\arcsec)}
   & \colhead{Type\tablenotemark{b}}
   & \colhead{(mag)}
   & \colhead{($\kms$)}
   & \colhead{(10$^{-17}$ erg s$^{-1}$ cm$^{-2}$)}
}
\startdata
1 &  1 & 23:23:28.54 ~ +58:50:47.6 & 8.75 &  S &  8.0 (0.5) & $+1698$ ( 2) &  589 ( 8) \\
1 &  2 & 23:23:28.54 ~ +58:50:34.1 & 3.25 &  S & 10.6 (0.7) & $ +225$ ( 2) &  165 ( 4) \\
1 &  3 & 23:23:28.54 ~ +58:50:33.8 & 5.00 & He &  5.8 (0.1) & $  -75$ ( 2) &  793 ( 5) \\
1 &  4 & 23:23:28.54 ~ +58:50:32.3 & 5.50 &  S &  3.3 (2.0) & $-2234$ (17) &   96 (13) \\
1 &  5 & 23:23:28.54 ~ +58:50:31.1 & 5.75 &  S &  4.9 (0.2) & $-1083$ ( 2) &  383 ( 3) \\
1 &  6 & 23:23:28.54 ~ +58:50:27.6 & 3.00 &  S &  3.3 (0.5) & $-1091$ ( 3) &  132 ( 4) \\
1 &  7 & 23:23:28.54 ~ +58:50:26.6 & 5.75 & Fe &  9.8 (0.4) & $  -17$ ( 2) &  561 (10) \\
1 &  8 & 23:23:28.54 ~ +58:50:25.8 & 3.00 &  S &  3.4 (0.4) & $-1264$ ( 2) &  198 ( 5) \\
2 &  1 & 23:23:35.19 ~ +58:50:06.0 & 4.50 & Fe &  5.5 (0.9) & $ +185$ ( 6) &  204 (10) \\
2 &  2 & 23:23:35.10 ~ +58:50:06.3 & 5.75 &  S &  9.3 (3.6) & $+1271$ (14) &   75 (10) \\
2 &  3 & 23:23:35.83 ~ +58:50:04.3 & 8.00 &  S &  4.2 (1.7) & $ -653$ (20) &  140 (12) \\
2 &  4 & 23:23:38.21 ~ +58:49:58.1 & 9.25 &  S &  7.0 (0.5) & $ +814$ ( 3) &  480 (10) \\
2 &  5 & 23:23:38.21 ~ +58:49:58.1 & 6.50 &  S & 11.0 (1.0) & $+1124$ ( 3) &  339 (12) \\
2 &  6 & 23:23:38.70 ~ +58:49:56.8 & 3.75 &  S &  9.2 (1.7) & $+1337$ ( 2) &  164 ( 7) \\
3 &  1 & 23:23:41.64 ~ +58:48:50.9 & 6.75 &  S &  8.2 (0.3) & $ -473$ ( 3) &  460 ( 6) \\
3 &  2 & 23:23:41.28 ~ +58:48:47.0 & 5.00 &  S &  7.6 (0.3) & $ -397$ ( 3) &  310 ( 4) \\
3 &  3 & 23:23:40.92 ~ +58:48:43.2 & 7.25 &  S &  7.3 (0.1) & $ -297$ ( 2) &  972 ( 4) \\
3 &  4 & 23:23:40.52 ~ +58:48:38.9 & 6.50 &  S &  7.2 (0.3) & $ +222$ ( 4) &  434 ( 8) \\
3 &  5 & 23:23:40.16 ~ +58:48:35.1 & 5.75 &  S &  6.2 (1.2) & $ +341$ ( 7) &  121 ( 7) \\
4 &  1 & 23:23:24.90 ~ +58:47:16.8 & 5.25 & Fe & 15.4 (1.4) & $+1109$ ( 5) &  240 ( 6) \\
4 &  2 & 23:23:24.82 ~ +58:47:13.9 & 3.75 &  S & 10.6 (0.6) & $+1306$ ( 4) &  289 ( 4) \\
4 &  3 & 23:23:24.77 ~ +58:47:12.2 & 5.75 & Fe &  7.5 (0.4) & $ -491$ ( 3) &  298 ( 5) \\
4 &  4 & 23:23:24.73 ~ +58:47:10.7 & 6.25 &  S &  9.9 (0.3) & $ +907$ ( 3) &  606 ( 5) \\
4 &  5 & 23:23:24.73 ~ +58:47:10.7 & 4.00 & Fe &  6.6 (0.4) & $  -31$ ( 4) &  236 ( 4) \\
4 &  6 & 23:23:24.72 ~ +58:47:10.2 & 5.75 &  S & 10.3 (0.3) & $ +390$ ( 3) &  577 ( 6) \\
5 &  1 & 23:23:17.21 ~ +58:47:32.7 & 5.25 &  S & 11.5 (0.8) & $+1924$ ( 3) &  313 ( 7) \\
5 &  2 & 23:23:17.28 ~ +58:47:32.3 & 5.25 &  S &  8.8 (0.5) & $ +992$ ( 3) &  561 (10) \\
5 &  3 & 23:23:17.34 ~ +58:47:32.0 & 4.50 & Fe &  8.7 (0.9) & $+1388$ ( 4) &  286 ( 8) \\
5 & 4A\tablenotemark{e} & 23:23:17.26 ~ +58:47:32.4 & 6.25 &  S &  8.3 (0.3) & $ -109$ ( 3) &  537 ( 7) \\
5 & 4B\tablenotemark{e} & 23:23:17.34 ~ +58:47:32.0 & 6.25 & He &  8.3 (0.1) & $ -109$ ( 3) & 2149 ( 7) \\
5 &  5 & 23:23:17.62 ~ +58:47:30.3 & 5.25 &  S &  9.3 (0.5) & $+1550$ ( 3) &  456 ( 6) \\
5 &  6 & 23:23:17.88 ~ +58:47:28.8 & 4.75 &  S &  8.8 (0.8) & $ -177$ ( 4) &  273 ( 8) \\
5 &  7 & 23:23:18.00 ~ +58:47:28.0 & 3.75 &  S &  7.2 (0.6) & $ -492$ ( 3) &  215 ( 5) \\
5 &  8 & 23:23:18.72 ~ +58:47:23.8 & 5.00 &  S & 10.8 (0.5) & $ +624$ ( 3) &  492 ( 9) \\
5 &  9 & 23:23:19.41 ~ +58:47:19.7 & 1.75 &  S & 11.9 (1.8) & $+1530$ ( 8) &   99 ( 5) \\
5 & 10 & 23:23:19.49 ~ +58:47:19.3 & 9.75 & Fe &  7.9 (0.0) & $ -290$ ( 3) & 4959 ( 8) \\
6 &  1 & 23:23:17.02 ~ +58:47:31.3 & 3.50 &  S &  6.6 (0.9) & $+1000$ ( 6) &  245 ( 8) \\
6 &  2 & 23:23:17.07 ~ +58:47:31.0 & 3.25 &  S &  9.0 (0.7) & $+1412$ ( 4) &  340 ( 7) \\
6 &  3 & 23:23:16.94 ~ +58:47:31.7 & 4.25 &  S & 10.2 (0.3) & $+1914$ ( 3) &  667 ( 7) \\
6 & 4A\tablenotemark{e} & 23:23:17.22 ~ +58:47:30.0 & 2.75 & He & 10.0 (2.1) & $ -166$ ( 6) &   58 ( 6) \\
6 & 4B\tablenotemark{e} & 23:23:17.30 ~ +58:47:29.6 & 2.75 &  S & 10.4 (3.6) & $ -166$ ( 6) &   34 ( 6) \\
6 &  5 & 23:23:17.22 ~ +58:47:30.0 & 3.00 &  S &  7.4 (0.8) & $+1343$ ( 5) &  272 ( 6) \\
6 &  6 & 23:23:17.53 ~ +58:47:28.2 & 6.25 &  S & 10.4 (0.4) & $+1547$ ( 3) &  714 ( 8) \\
6 &  7 & 23:23:17.99 ~ +58:47:25.5 & 4.50 &  S &  7.8 (0.3) & $ +802$ ( 3) &  623 ( 8) \\
6 &  8 & 23:23:18.47 ~ +58:47:22.6 & 2.75 &  S &  9.3 (0.7) & $ +572$ ( 4) &  171 ( 5) \\
6 &  9 & 23:23:18.75 ~ +58:47:21.0 & 2.75 & Fe &  8.3 (0.5) & $ +540$ ( 4) &  258 ( 5) \\
6 & 10 & 23:23:19.04 ~ +58:47:19.3 & 3.50 & Fe &  7.3 (0.4) & $ +459$ ( 3) &  283 ( 5) \\
6 & 11 & 23:23:19.29 ~ +58:47:17.8 & 6.50 & Fe &  6.9 (0.2) & $ -291$ ( 3) &  676 ( 7) \\
6 & 12 & 23:23:19.52 ~ +58:47:16.4 & 4.25 & Fe &  9.9 (1.7) & $ +293$ ( 6) &   92 ( 6) \\
7 &  1 & 23:23:13.66 ~ +58:47:56.4 & 3.25 &  S & 10.2 (0.7) & $+1591$ ( 3) &  161 ( 3) \\
7 &  2 & 23:23:13.62 ~ +58:47:53.2 & 5.50 & He &  8.4 (0.2) & $  -83$ ( 3) &  424 ( 6) \\
7 &  3 & 23:23:13.51 ~ +58:47:44.2 & 4.00 & He &  8.5 (0.2) & $  +42$ ( 2) &  647 ( 9) \\
7 &  4 & 23:23:13.49 ~ +58:47:42.0 & 2.75 & He &  8.5 (0.3) & $  +78$ ( 3) &  480 (10) \\
7 &  5 & 23:23:13.46 ~ +58:47:39.7 & 3.25 & He &  9.1 (0.2) & $  -63$ ( 3) &  516 ( 7) \\
8 &  1 & 23:23:18.56 ~ +58:50:28.1 & 3.25 &  S &  8.1 (0.9) & $ +727$ ( 3) &  180 ( 7) \\
8 &  2 & 23:23:18.43 ~ +58:50:24.0 & 6.00 &  S &  7.6 (0.5) & $ +841$ ( 3) &  369 ( 9) \\
8 &  3 & 23:23:18.22 ~ +58:50:17.1 & 6.25 &  S &  7.3 (0.4) & $ +737$ ( 2) &  443 ( 9) \\
8 &  4 & 23:23:18.07 ~ +58:50:12.3 & 5.25 &  S &  9.2 (0.4) & $ +621$ ( 3) &  364 ( 6) \\
8 &  5 & 23:23:17.93 ~ +58:50:07.7 & 4.50 &  S & 10.8 (1.6) & $ +606$ ( 6) &  170 ( 8) \\
8 &  6 & 23:23:17.88 ~ +58:50:06.2 & 4.00 &  S & 13.6 (1.7) & $ +478$ ( 3) &  139 ( 5) \\
8 &  7 & 23:23:17.76 ~ +58:50:02.3 & 4.75 &  S &  9.4 (0.7) & $ +627$ ( 4) &  378 (10) \\
8 &  8 & 23:23:17.76 ~ +58:50:02.1 & 4.00 &  S &  6.8 (0.8) & $ +146$ ( 5) &  239 (12) \\
8 &  9 & 23:23:17.67 ~ +58:49:59.1 & 4.00 &  S &  8.1 (0.6) & $ +361$ ( 3) &  294 ( 6) \\
\enddata
\tablenotetext{a}{
Size along the slit length.
}
\tablenotetext{b}{
He: He-rich knot, S: S-rich knot, Fe: Fe-rich knot
}
\tablenotetext{c}{
Visual extinction derived from the flux ratio of [Fe II] 1.257 and 1.644 \micron.
We adopted the intrinsic [Fe II] line ratio of 1.36 \citep{deb10}
and the extinction curve of the Milky Way with $R_{\rm V} = 3.1$
\citep[][see Section~\ref{sec-pca-method} for more details]{dra03}.
}
\tablenotetext{d}{
The uncertainty in parenthesis is $1\sigma$ statistical error by a single Gaussian fitting,
and does not include the absolute photometric error which is roughly 20\% or less.
}
\tablenotetext{e}{
Knot 4 in Slits 5 and 6 have been identified as a single knot by {\sc Clumpfind} respectively,
but a detailed inspection revealed that each of them are composed of two (A and B) components
almost coincident both in space and velocity.
}
\end{deluxetable}

\clearpage
\begin{deluxetable}{cc|lr|rr|r}
\tablewidth{0pt}
\tabletypesize{\tiny}
\tablecaption{Observed NIR Line Parameters of Knots \label{tbl-flx}}
\tablehead {
\colhead{Slit} & \colhead{Knot} & \colhead{Line ID}
   & \colhead{$\lambda_{\rm rest}$\tablenotemark{b}}
   & \colhead{${\rm FWHM}$\tablenotemark{c}}
   & \colhead{Observed Flux\tablenotemark{d}}
   & \colhead{Note\tablenotemark{e}} \\
\colhead{No.} & \colhead{No.\tablenotemark{a}} & \colhead{Transition (l - u)}
   & \colhead{(\micron)}
   & \colhead{(\AA)}
   & \colhead{(10$^{-17}$ erg s$^{-1}$ cm$^{-2}$)}
   & \colhead{ } }
\startdata
1 &  1 & [S III] ~ $^{3}P_{2}$~-~$^{1}D_{2}$            & 0.95311 &    7.2   (     0.2    ) &   9233   ( 261  ) &  \\
1 &  1 & [C I] ~~~ $^{3}P_{1}$~-~$^{1}D_{2}$            & 0.98241 & $\cdot$~ (~~~$\cdot$~~) & $\cdot$~ (~~15~~) &  \\
1 &  1 & [C I] ~~~ $^{3}P_{2}$~-~$^{1}D_{2}$            & 0.98503 & $\cdot$~ (~~~$\cdot$~~) & $\cdot$~ (~~15~~) &  \\
1 &  1 & [S II] ~~ $^{2}D_{3/2}$~-~$^{2}P_{3/2}$        & 1.02867 &    8.4   (     0.1    ) &   2317   (~~18~~) & LINE-FIX \\
1 &  1 & [S II] ~~ $^{2}D_{5/2}$~-~$^{2}P_{3/2}$        & 1.03205 &    8.4   (     ~-~    ) &   3182   ( ~~-~~) & LINE-FIX \\
1 &  1 & [S II] ~~ $^{2}D_{3/2}$~-~$^{2}P_{1/2}$        & 1.03364 &    8.4   (     ~-~    ) &   2185   (~~13~~) & LINE-FIX \\
1 &  1 & [S II] ~~ $^{2}D_{5/2}$~-~$^{2}P_{1/2}$        & 1.03705 &    8.4   (     ~-~    ) &   1058   ( ~~-~~) & LINE-FIX \\
1 &  1 & [N I] ~~~ $^{2}D_{5/2}$~-~$^{2}P_{3/2,1/2}$    & 1.03979 & $\cdot$~ (~~~$\cdot$~~) & $\cdot$~ (~~12~~) &  \\
1 &  1 & [N I] ~~~ $^{2}D_{3/2}$~-~$^{2}P_{3/2,1/2}$    & 1.04074 & $\cdot$~ (~~~$\cdot$~~) & $\cdot$~ (~~14~~) &  \\
1 &  1 & [S I] ~~~ $^{3}P_{2}$~-~$^{1}D_{2}$            & 1.08212 &    9.9   (     0.5    ) &    207   (~~15~~) &  \\
1 &  1 & He I ~~~~ $^{3}S_{1}$~-~$^{3}P_{0,1,2}$        & 1.08302 & $\cdot$~ (~~~$\cdot$~~) & $\cdot$~ (~~11~~) &  \\
1 &  1 & H I Pa$\gamma$                                 & 1.09381 & $\cdot$~ (~~~$\cdot$~~) & $\cdot$~ (~~~9~~) &  \\
1 &  1 & [S I] ~~~ $^{3}P_{1}$~-~$^{1}D_{2}$            & 1.13059 & $\cdot$~ (~~~$\cdot$~~) & $\cdot$~ (~~13~~) &  \\
1 &  1 & [P II] ~~ $^{3}P_{1}$~-~$^{1}D_{2}$            & 1.14682 &    7.8   (     0.7    ) &    293   (~~48~~) & OH-CONT \\
1 &  1 & [P II] ~~ $^{3}P_{2}$~-~$^{1}D_{2}$            & 1.18828 &    9.8   (     0.1    ) &    736   (~~15~~) &  \\
1 &  1 & [Fe II] ~ $a^{6}D_{7/2}$~-~$a^{4}D_{5/2}$      & 1.24854 & $\cdot$~ (~~~$\cdot$~~) & $\cdot$~ (~~~7~~) &  \\
1 &  1 & [Fe II] ~ $a^{6}D_{3/2}$~-~$a^{4}D_{1/2}$      & 1.25214 & $\cdot$~ (~~~$\cdot$~~) & $\cdot$~ (~~~8~~) &  \\
1 &  1 & [Fe II] ~ $a^{6}D_{9/2}$~-~$a^{4}D_{7/2}$      & 1.25668 &   10.2   (     0.3    ) &    388   (~~15~~) &  \\
1 &  1 & [Fe II] ~ $a^{6}D_{1/2}$~-~$a^{4}D_{1/2}$      & 1.27035 & $\cdot$~ (~~~$\cdot$~~) & $\cdot$~ (~~18~~) &  \\
1 &  1 & [Fe II] ~ $a^{6}D_{3/2}$~-~$a^{4}D_{3/2}$      & 1.27878 &    4.0   (     0.6    ) &     30   (~~~6~~) &  \\
1 &  1 & H I Pa$\beta$                                  & 1.28181 & $\cdot$~ (~~~$\cdot$~~) & $\cdot$~ (~~12~~) &  \\
1 &  1 & [Fe II] ~ $a^{6}D_{5/2}$~-~$a^{4}D_{5/2}$      & 1.29427 &    9.3   (     1.0    ) &    133   (~~22~~) & OH-CONT \\
1 &  1 & [Fe II] ~ $a^{6}D_{1/2}$~-~$a^{4}D_{3/2}$      & 1.29777 & $\cdot$~ (~~~$\cdot$~~) & $\cdot$~ (~~12~~) &  \\
1 &  1 & [Fe II] ~ $a^{6}D_{7/2}$~-~$a^{4}D_{7/2}$      & 1.32055 &   17.5   (     2.6    ) &     92   (~~18~~) &  \\
1 &  1 & [Fe II] ~ $a^{6}D_{3/2}$~-~$a^{4}D_{5/2}$      & 1.32778 &   12.3   (     2.1    ) &     67   (~~15~~) &  \\
1 &  1 & [Fe II] ~ $a^{4}F_{9/2}$~-~$a^{4}D_{5/2}$      & 1.53347 &    8.7   (     0.9    ) &     95   (~~12~~) & OH-CONT \\
1 &  1 & [Fe II] ~ $a^{4}F_{7/2}$~-~$a^{4}D_{3/2}$      & 1.59947 &    9.8   (     1.1    ) &     76   (~~11~~) & OH-CONT \\
1 &  1 & [Si I] ~~ $^{3}P_{1}$~-~$^{1}D_{2}$            & 1.60683 & $\cdot$~ (~~~$\cdot$~~) & $\cdot$~ (~~~5~~) &  \\
1 &  1 & [Fe II] ~ $a^{4}F_{9/2}$~-~$a^{4}D_{7/2}$      & 1.64355 &   13.7   (     0.1    ) &    589   (~~~8~~) &  \\
1 &  1 & [Si I] ~~ $^{3}P_{2}$~-~$^{1}D_{2}$            & 1.64545 & $\cdot$~ (~~~$\cdot$~~) & $\cdot$~ (~~~7~~) &  \\
1 &  1 & [Fe II] ~ $a^{4}F_{5/2}$~-~$a^{4}D_{1/2}$      & 1.66377 &   14.0   (     1.5    ) &     73   (~~11~~) & OH-CONT \\
1 &  1 & [Fe II] ~ $a^{4}F_{7/2}$~-~$a^{4}D_{5/2}$      & 1.67688 &   15.4   (     0.5    ) &    146   (~~~8~~) &  \\
1 &  1 & [Fe II] ~ $a^{4}F_{5/2}$~-~$a^{4}D_{3/2}$      & 1.71113 & $\cdot$~ (~~~$\cdot$~~) & $\cdot$~ (~~~9~~) &  \\
1 &  1 & [Fe II] ~ $a^{4}F_{3/2}$~-~$a^{4}D_{1/2}$      & 1.74494 & $\cdot$~ (~~~$\cdot$~~) & $\cdot$~ (~~~5~~) &  \\
1 &  1 & [Fe II] ~ $a^{4}F_{3/2}$~-~$a^{4}D_{3/2}$      & 1.79710 & $\cdot$~ (~~~$\cdot$~~) & $\cdot$~ (~~62~~) &  \\
1 &  1 & [Fe II] ~ $a^{4}F_{5/2}$~-~$a^{4}D_{5/2}$      & 1.80002 & $\cdot$~ (~~~$\cdot$~~) & $\cdot$~ ( 240  ) &  \\
1 &  1 & [Fe II] ~ $a^{4}F_{7/2}$~-~$a^{4}D_{7/2}$      & 1.80939 & $\cdot$~ (~~~$\cdot$~~) & $\cdot$~ ( 689  ) &  \\
1 &  1 & [Si VI] ~ $^{2}P_{3/2}$~-~$^{2}P_{1/2}$        & 1.96287 &   16.4   (     0.6    ) &    501   (~~27~~) &  \\
1 &  1 & [Fe II] ~ $a^{4}P_{5/2}$~-~$a^{2}P_{3/2}$      & 2.04601 & $\cdot$~ (~~~$\cdot$~~) & $\cdot$~ (~~20~~) &  \\
1 &  1 &  He I ~~~ $^{1}S_{0}$~-~$^{1}P_{1}$            & 2.05813 & $\cdot$~ (~~~$\cdot$~~) & $\cdot$~ (~~12~~) &  \\
1 &  1 & [Fe II] ~ $a^{4}P_{3/2}$~-~$a^{2}P_{3/2}$      & 2.13277 & $\cdot$~ (~~~$\cdot$~~) & $\cdot$~ (~~~9~~) &  \\
1 &  1 & [Fe III]~ $a^{3}H_{4}$~-~$a^{3}G_{3}$          & 2.14511 & $\cdot$~ (~~~$\cdot$~~) & $\cdot$~ (~~10~~) &  \\
1 &  1 & H I Br$\gamma$                                 & 2.16553 & $\cdot$~ (~~~$\cdot$~~) & $\cdot$~ (~~11~~) &  \\
1 &  1 & [Fe III]~ $a^{3}H_{6}$~-~$a^{3}G_{5}$          & 2.21779 & $\cdot$~ (~~~$\cdot$~~) & $\cdot$~ (~~13~~) &  \\
1 &  1 & [Fe II] ~ $a^{4}G_{9/2}$~-~$a^{2}H_{11/2}$     & 2.22379 & $\cdot$~ (~~~$\cdot$~~) & $\cdot$~ (~~10~~) &  \\
1 &  1 & [Fe III]~ $a^{3}H_{4}$~-~$a^{3}G_{4}$          & 2.24209 & $\cdot$~ (~~~$\cdot$~~) & $\cdot$~ (~~13~~) &  \\
\enddata
\tablecomments{
This table is available in its entirety in a machine-readable form
in the online journal.
}
\tablenotetext{a}{
Knot 4 in Slits 5 and 6 has been identified as a single knot by {\sc Clumpfind},
but a detailed inspection revealed that each of them are composed of two (A and B) components
almost coincident both in space and velocity.
}
\tablenotetext{b}{
Rest wavelengths in air.
}
\tablenotetext{c}{
FWHM of lines.
}
\tablenotetext{d}{
The uncertainty in parenthesis is the $1\sigma$ statistical error from a single Gaussian fitting
and does not include the absolute photometric error which is roughly 20\% or less.
The uncertainty of the undetected lines was derived from
the background rms noise around the wavelength.
The emission lines falling in the bad atmospheric transmission window
(e.g., [Fe II] lines near 1.80~\micron)
have much higher uncertainty in flux due to low signal-to-noise ratios.
}
\tablenotetext{e}{
In the case of the lines which were contaminated by nearby lines
of similar wavelengths either from the knot itself or from other knots,
we carried out a simultaneous Gaussian fitting
with possible constraints (``LINE-FIX'' keyword in Note),
e.g., by fixing their wavelengths and/or line widths
based on the parameters of well-isolated lines,
by fixing their intensities if they can be predictable theoretically
(\citet{fro06} for [C I],  \citet{kel08} for [Si I], \citet{tay10} for [S II],
\citet{deb10} for [Fe II]).
``OH-CONT'' keyword in Note represents the line
which was significantly contaminated by
nearby bright OH airglow emission lines
so their true uncertainty could be much larger than
what we measured from a single Gaussian fitting.
}
\end{deluxetable}

\clearpage
\begin{deluxetable}{ccrr}
\tablewidth{0pt}
\tablecaption{Individual and Cumulative Fraction of Variance \label{tbl-pca}}
\tablehead {
\colhead{Number}	& \colhead{Eigenvalue}	& \colhead{Fraction of}		& \colhead{Cumulative} \\
\colhead{of PC}		& \colhead{ }			& \colhead{Variance [\%]}	& \colhead{Fraction [\%]} }
\startdata
 1     &    9.8    &    42.7    &    42.7 \\
 2     &    5.1    &    22.2    &    64.9 \\
 3     &    4.7    &    20.4    &    85.3 \\
 4     &    1.1    &     4.7    &    90.0 \\
 5     &    1.0    &     4.2    &    94.2 \\
 6     &    0.4    &     1.8    &    96.0 \\
 7     &    0.3    &     1.3    &    97.4 \\
 8     &    0.2    &     0.7    &    98.1 \\
 9     &    0.1    &     0.5    &    98.6 \\
10     &    0.1    &     0.4    &    98.9 \\
\enddata
\end{deluxetable}


\begin{thebibliography}{}

\bibitem[Alarie et al.(2014)]{ala14} Alarie, A., Bilodeau, A., \& Drissen, L.\ 2014, \mnras, 441, 2996 

\bibitem[Barlow et al.(2010)]{bar10} Barlow, M.~J., Krause, O., Swinyard, B.~M., et al.\ 2010, \aap, 518, L138 

\bibitem[Blondin et al.(2001)]{blo01} Blondin, J.~M., Borkowski, K.~J., \& Reynolds, S.~P.\ 2001, \apj, 557, 782

\bibitem[Chevalier(1982)]{che82} Chevalier, R.~A.\ 1982, \apj, 258, 790

\bibitem[Chevalier et al.(1992)]{che92} Chevalier, R.~A., Blondin, J.~M., \& Emmering, R.~T.\ 1992, \apj, 392, 118 

\bibitem[Chevalier \& Kirshner(1979)]{che79} Chevalier, R.~A., \& Kirshner, R.~P.\ 1979, \apj, 233, 154 

\bibitem[Chevalier \& Oishi(2003)]{che03} Chevalier, R.~A., \& Oishi, J.\ 2003, \apjl, 593, L23

\bibitem[Deb \& Hibbert(2010)]{deb10} Deb, N.~C., \& Hibbert, A.\ 2010, \apjl, 711, L104 

\bibitem[DeLaney et al.(2010)]{del10} DeLaney, T., Rudnick, L., Stage, M.~D., et al.\ 2010, \apj, 725, 2038 

\bibitem[Dennefeld \& Andrillat(1981)]{den81} Dennefeld, M., \& Andrillat, Y.\ 1981, \aap, 103, 44

\bibitem[Draine(2003)]{dra03} Draine, B.~T.\ 2003, \apj, 598, 1017 

\bibitem[Ennis et al.(2006)]{enn06} Ennis, J.~A., Rudnick, L., Reach, W.~T., et al.\ 2006, \apj, 652, 376 

\bibitem[Eriksen et al.(2009)]{eri09} Eriksen, K.~A., Arnett, D., McCarthy, D.~W., \& Young, P.\ 2009, \apj, 697, 29 

\bibitem[Fesen(2001)]{fes01a} Fesen, R.~A.\ 2001, \apjs, 133, 161 

\bibitem[Fesen et al.(1987)]{fes87} Fesen, R.~A., Becker, R.~H., \& Blair, W.~P.\ 1987, \apj, 313, 378 

\bibitem[Fesen et al.(1988)]{fes88} Fesen, R.~A., Becker, R.~H., \& Goodrich, R.~W.\ 1988, \apjl, 329, L89

\bibitem[Fesen \& Gunderson(1996)]{fes96} Fesen, R.~A., \& Gunderson, K.~S.\ 1996, \apj, 470, 967 

\bibitem[Fesen et al.(2006)]{fes06a} Fesen, R.~A., Hammell, M.~C., Morse, J., et al.\ 2006, \apj, 636, 859 

\bibitem[Fesen et al.(2006)]{fes06b} Fesen, R.~A., Hammell, M.~C., Morse, J., et al.\ 2006, \apj, 645, 283 

\bibitem[Fesen et al.(2001)]{fes01b} Fesen, R.~A., Morse, J.~A., Chevalier, R.~A., et al.\ 2001, \aj, 122, 2644

\bibitem[Froese Fischer(2006)]{fro06} Froese Fischer, C.\ 2006, Journal of Physics B Atomic Molecular Physics, 39, 2159

\bibitem[Gerardy \& Fesen(2001)]{ger01} Gerardy, C.~L., \& Fesen, R.~A.\ 2001, \aj, 121, 2781 

\bibitem[Giannini et al.(2015)]{gia15} Giannini, T., Antoniucci, S., Nisini, B., et al.\ 2015, \apj, 798, 33 

\bibitem[Gilkis \& Soker(2015)]{gil15} Gilkis, A., \& Soker, N.\ 2015, \apj, 806, 28 

\bibitem[Grefenstette et al.(2014)]{gre14} Grefenstette, B.~W., Harrison, F.~A., Boggs, S.~E., et al.\ 2014, \nat, 506, 339

\bibitem[Hammell \& Fesen(2008)]{ham08} Hammell, M.~C., \& Fesen, R.~A.\ 2008, \apjs, 179, 195 

\bibitem[Hammer et al.(2010)]{ham10} Hammer, N.~J., Janka, H.-T., M{\"u}ller, E.\ 2010, \apj, 714, 1371

\bibitem[Herter et al.(2008)]{her08} Herter, T.~L., Henderson, C.~P., Wilson, J.~C., et al.\ 2008, \procspie, 7014, 70140X 

\bibitem[Hughes et al.(2000)]{hug00} Hughes, J.~P., Rakowski, C.~E., Burrows, D.~N., \& Slane, P.~O.\ 2000, \apjl, 528, L109 

\bibitem[Hurford \& Fesen(1996)]{hur96} Hurford, A.~P., \& Fesen, R.~A.\ 1996, \apj, 469, 246 

\bibitem[Hwang \& Laming(2003)]{hwa03} Hwang, U., \& Laming, J.~M.\ 2003, \apj, 597, 362 

\bibitem[Hwang \& Laming(2012)]{hwa12} Hwang, U., \& Laming, J.~M.\ 2012, \apj, 746, 130 

\bibitem[Hwang et al.(2004)]{hwa04} Hwang, U., Laming, J.~M., Badenes, C., et al.\ 2004, \apjl, 615, L117 

\bibitem[Isensee et al.(2010)]{ise10} Isensee, K., Rudnick, L., DeLaney, T., et al.\ 2010, \apj, 725, 2059

\bibitem[Joggerst et al.(2009)]{jog09} Joggerst, C.~C., Woosley, S.~E., \& Heger, A.\ 2009, \apj, 693, 1780 

\bibitem[Kelleher \& Podobedova(2008)]{kel08} Kelleher, D.~E., \& Podobedova, L.~I.\ 2008, Journal of Physical and Chemical Reference Data, 37, 1285

\bibitem[Kifonidis et al.(2003)]{kif03} Kifonidis, K., Plewa, T., Janka, H.-T., M{\"u}ller, E.\ 2003, \aap, 408, 621 

\bibitem[Kifonidis et al.(2006)]{kif06} Kifonidis, K., Plewa, T., Scheck, L., Janka, H.-T., M{\"u}ller, E.\ 2006, \aap, 453, 661 

\bibitem[Koo \& Lee(2015)]{koo15} Koo, B.-C., \& Lee, Y.-H.\ 2015, Publication of Korean Astronomical Society, 30, 145 

\bibitem[Koo et al.(2013)]{koo13} Koo, B.-C., Lee, Y.-H., Moon, D.-S., Yoon, S.-C., \& Raymond, J.~C.\ 2013, Science, 342, 1346

\bibitem[Koo et al.(2016)]{koo16} Koo, B.-C., Raymond, J.~C., \& Kim, H.-J.\ 2016, Journal of Korean Astronomical Society, 49, 109

\bibitem[Krause et al.(2008)]{kra08} Krause, O., Birkmann, S.~M., Usuda, T., et al.\ 2008, Science, 320, 1195 

\bibitem[Lee et al.(2014)]{leejj14} Lee, J.-J., Park, S., Hughes, J.~P., \& Slane, P.~O.\ 2014, \apj, 789, 7 

\bibitem[Lee et al.(2015)]{leeyh15} Lee, Y.-H., Koo, B.-C., Moon, D.-S., \& Lee, J.-J.\ 2015, \apj, 808, 98 

\bibitem[Likkel et al.(2006)]{lik06} Likkel, L., Dinerstein, H.~L., Lester, D.~F., Kindt, A., \& Bartig, K.\ 2006, \aj, 131, 1515 

\bibitem[Lutz et al.(1993)]{lut93} Lutz, D., Krabbe, A., \& Genzel, R.\ 1993, \apj, 418, 244 

\bibitem[Magkotsios et al.(2010)]{mag10} Magkotsios, G., Timmes, F.~X., Hungerford, A.~L., et al.\ 2010, \apjs, 191, 66 

\bibitem[Mao et al.(2015)]{mao15} Mao, J., Ono, M., Nagataki, S., et al.\ 2015, \apj, 808, 164 

\bibitem[Milisavljevic \& Fesen(2013)]{mil13} Milisavljevic, D., \& Fesen, R.~A.\ 2013, \apj, 772, 134 

\bibitem[Moon et al.(2009)]{moo09} Moon, D.-S., Koo, B.-C., Lee, H.-G., et al.\ 2009, \apjl, 703, L81 

\bibitem[Neufeld et al.(2007)]{neu07} Neufeld, D.~A., Hollenbach, D.~J., Kaufman, M.~J., et al.\ 2007, \apj, 664, 890 

\bibitem[Nussbaumer \& Storey(1988)]{nus88} Nussbaumer, H., \& Storey, P.~J.\ 1988, \aap, 193, 327 

\bibitem[Okumura et al.(2001)]{oku01} Okumura, S.-i., Mori, A., Watanabe, E., Nishihara, E., \& Yamashita, T.\ 2001, \aj, 121, 2089 

\bibitem[Osterbrock et al.(1997)]{ost97} Osterbrock, D.~E., Fulbright, J.~P., \& Bida, T.~A.\ 1997, \pasp, 109, 614

\bibitem[Reed et al.(1995)]{ree95} Reed, J.~E., Hester, J.~J., Fabian, A.~C., \& Winkler, P.~F.\ 1995, \apj, 440, 706 

\bibitem[Rest et al.(2011)]{res11} Rest, A., Foley, R.~J., Sinnott, B., et al.\ 2011, \apj, 732, 3 

\bibitem[Rho et al.(2003)]{rho03} Rho, J., Reynolds, S.~P., Reach, W.~T., et al.\ 2003, \apj, 592, 299 

\bibitem[Rousselot et al.(2000)]{rou00} Rousselot, P., Lidman, C., Cuby, J.-G., Moreels, G., \& Monnet, G.\ 2000, \aap, 354, 1134 

\bibitem[Rudy et al.(1994)]{rud94} Rudy, R.~J., Rossano, G.~S., \& Puetter, R.~C.\ 1994, \apj, 426, 646 

\bibitem[Sibthorpe et al.(2010)]{sib10} Sibthorpe, B., Ade, P.~A.~R., Bock, J.~J., et al.\ 2010, \apj, 719, 1553 

\bibitem[Smith et al.(2009)]{smi09} Smith, J.~D.~T., Rudnick, L., Delaney, T., et al.\ 2009, \apj, 693, 713 

\bibitem[Skrutskie et al.(2006)]{skr06} Skrutskie, M.~F., Cutri, R.~M., Stiening, R., et al.\ 2006, \aj, 131, 1163 

\bibitem[Sumiyoshi et al.(2005)]{sum05} Sumiyoshi, K., Yamada, S., Suzuki, H., et al.\ 2005, \apj, 629, 922 

\bibitem[Takiwaki et al.(2014)]{tak14} Takiwaki, T., Kotake, K., \& Suwa, Y.\ 2014, \apj, 786, 83 

\bibitem[Tayal \& Zatsarinny(2010)]{tay10} Tayal, S.~S., \& Zatsarinny, O.\ 2010, \apjs, 188, 32

\bibitem[Thorstensen et al.(2001)]{tho01} Thorstensen, J.~R., Fesen, R.~A., \& van den Bergh, S.\ 2001, \aj, 122, 297

\bibitem[Ungerechts et al.(1997)]{ung97} Ungerechts, H., Bergin, E.~A., Goldsmith, P.~F., et al.\ 1997, \apj, 482, 245

\bibitem[van den Bergh \& Kamper(1985)]{van85} van den Bergh, S., \& Kamper, K.\ 1985, \apj, 293, 537

\bibitem[van Veelen et al.(2009)]{van09} van Veelen, B., Langer, N., Vink, J., Garc{\'{\i}}a-Segura, G., \& van Marle, A.~J.\ 2009, \aap, 503, 495 

\bibitem[Wagner \& Depoy(1996)]{wag96} Wagner, R.~M., \& Depoy, D.~L.\ 1996, \apj, 467, 860 

\bibitem[Williams et al.(1994)]{wil94} Williams, J.~P., de Geus, E.~J., \& Blitz, L.\ 1994, \apj, 428, 693 

\bibitem[Wilson et al.(2003)]{wil03} Wilson, J.~C., Eikenberry, S.~S., Henderson, C.~P., et al.\ 2003, \procspie, 4841, 451 

\bibitem[Wilson et al.(2004)]{wil04} Wilson, J.~C., Henderson, C.~P., Herter, T.~L., et al.\ 2004, \procspie, 5492, 1295

\bibitem[Wongwathanarat et al.(2013)]{won13} Wongwathanarat, A., Janka, H.-T., \& M{\"u}ller, E.\ 2013, \aap, 552, A126 

\bibitem[Woosley et al.(1973)]{woo73} Woosley, S.~E., Arnett, W.~D., \& Clayton, D.~D.\ 1973, \apjs, 26, 231 

\end{thebibliography}
\end{document}